\documentclass[twocolumn,fleqn]{aastex631}
\usepackage{graphicx}	
\usepackage{mathrsfs}
\usepackage{amsmath}	
\usepackage{bm}
\usepackage{tikz}
\usepackage{amssymb}	
\usepackage{enumitem}   
\usepackage{soul} 
\usepackage[hang]{footmisc}
\newcommand{\SatGen}{{\tt SatGen}\,}

\newcommand{\eq}[1]{eq.~(\ref{eq:#1})}
\newcommand{\eqs}[1]{eqs.~(\ref{eq:#1})}

\newcommand{\se}[1]{Section \ref{sec:#1}}
\newcommand{\app}[1]{Appendix \ref{app:#1}}
\newcommand{\Fig}[1]{Fig.~\ref{fig:#1}}
\newcommand{\Figs}[1]{Figs.~\ref{fig:#1}}

\newcommand{\be}{\begin{equation}}
\newcommand{\ee}{\end{equation}}
\newcommand{\bad}{\begin{equation} \begin{aligned}}
\newcommand{\ead}{\end{aligned} \end{equation}}

\newcommand{\Msun}{M_\odot}

\newcommand{\kpc}{\,{\rm kpc}}

\newcommand{\Gyr}{\,{\rm Gyr}}

\newcommand{\g}{\,{\rm g}}
\newcommand{\kms}{\,{\rm km/s}}

\newcommand{\rhoc}{\rho_{\rm crit}}

\newcommand{\rhohalf}{\rho_{\rm 1/2}}

\newcommand{\taudyn}{\tau_{\rm dyn}}

\newcommand{\Mv}{M_{\rm vir}}
\newcommand{\Ms}{M_{\star}}
\newcommand{\Mh}{M_{\rm h}}

\newcommand{\Mmax}{M_{\rm max}}
\newcommand{\Mmin}{M_{\rm min}}

\newcommand{\mmax}{m_{\rm max}}
\newcommand{\mmin}{m_{\rm min}}

\newcommand{\mt}{m_{\rm t}}
\newcommand{\mr}{m_{\rm r}}

\newcommand{\fr}{f_{\rm r}}
\newcommand{\ft}{f_{\rm t}}
\newcommand{\Er}{E_{\rm r}}
\newcommand{\Et}{E_{\rm t}}

\newcommand{\rv}{r_{\rm vir}}

\newcommand{\rs}{r_{\rm s}}

\newcommand{\lt}{l_{\rm t}}

\newcommand{\lhalf}{l_{\rm 1/2}}
\newcommand{\lmax}{l_{\rm max}}
\newcommand{\bmax}{b_{\rm max}}
\newcommand{\bmin}{b_{\rm min}}
\newcommand{\Vc}{V_{\rm circ}}

\newcommand{\Vt}{V_{\rm t}}

\newcommand{\vmax}{v_{\rm max}}

\newcommand{\lnL}{\ln\Lambda}

\newcommand{\xit}{\xi_{\rm t}}

\newcommand{\rmd}{{\rm d}}
\newcommand{\dd}{\text{d}}

\begin{document}
\nolinenumbers

\title{Constrain the Dark Matter Distribution of Ultra-diffuse Galaxies with Globular-Cluster Mass Segregation: A Case Study with NGC5846-UDG1}

\author[0000-0001-8405-2921]{Jinning Liang}
\affiliation{Kavli Institute for Astronomy and Astrophysics, Peking University, Beijing 100871, China}
\affiliation{School of Physics and Technology, Wuhan University, Wuhan, Hubei 430072, China}

\author[0000-0001-6115-0633]{Fangzhou Jiang}\thanks{Corresponding author: \href{mailto:fangzhou.jiang@pku.edu.cn}{fangzhou.jiang@pku.edu.cn}}
\affiliation{Kavli Institute for Astronomy and Astrophysics, Peking University, Beijing 100871, China}
\affiliation{Carnegie Observatories, 813 Santa Barbara Street, Pasadena, CA 91101, USA}
\affiliation{TAPIR, California Institute of Technology, Pasadena, CA 91125, USA}

\author[0000-0002-1841-2252]{Shany Danieli}
\affiliation{Department of Astrophysical Sciences, 4 Ivy Lane, Princeton University, Princeton, NJ 08544}

\author[0000-0001-5501-6008]{Andrew Benson}
\affiliation{Carnegie Observatories, 813 Santa Barbara Street, Pasadena, CA 91101, USA}

\author[0000-0003-3729-1684]{Phil Hopkins}
\affiliation{TAPIR, California Institute of Technology, Pasadena, CA 91125, USA}



\begin{abstract}
The properties of globular clusters (GCs) contain valuable information of their host galaxies and dark-matter halos. In the remarkable example of ultra-diffuse galaxy, NGC5846-UDG1, the GC population exhibits strong radial mass segregation, indicative of dynamical-friction-driven orbital decay, which opens the possibility of using imaging data alone to constrain the dark-matter content of the galaxy. 
To explore this possibility, we develop a semi-analytical model of GC evolution, which starts from the initial mass, structural, and spatial distributions of the GC progenitors, and follows the effects of dynamical friction, tidal evolution, and two-body relaxation. Using Markov Chain Monte Carlo, we forward-model the GCs in a UDG1-like potential to match the observed GC statistics, and to constrain the profile of the host halo and the origin of the GCs. We find that, with the assumptions of zero mass segregation when the star clusters were born, UDG1 is relatively dark-matter poor compared to what is expected from stellar-to-halo-mass relations, and its halo concentration is lower than the cosmological average, irrespective of having a cuspy or a cored profile. Its GC population has an initial spatial distribution more extended than the smooth stellar distribution. We discuss the results in the context of scaling laws of galaxy-halo connections, and warn against naively using the GC-abundance-halo-mass relation to infer the halo mass of UDGs.  Our model is generally applicable to GC-rich dwarf galaxies, and
is publicly available at \href{https://github.com/JiangFangzhou/GCevo}{https://github.com/JiangFangzhou/GCevo}.
\end{abstract}



\section{Introduction} \label{sec:intro}
Ultra-diffuse galaxies (UDGs) triggered a frenzy of studies in recent years in the contexts of both understanding the formation of extreme galaxies and testing cosmology \citep[see e.g., the review by][and the references therein]{Sales20}.
Numerical and semi-analytical simulations suggest that UDGs can form via supernovae-driven gas outflows, which transform their hosting dark-matter halos from cuspy to cored together with puffing up their stellar distribution \citep[e.g.,][]{DiCintio17,Chan19,Jiang19}, while some UDGs can also populate halos of high specific angular momentum \citep[e.g.,][]{Rong17,Amorisco16,Benavides22}.
The formation of UDGs is also suggested to be facilitated in dense environments, via tidal heating \citep{Jiang19, Carleton19, Liao19} and passive stellar-population dimming \citep{Tremmel20}. 
Despite all these theoretical efforts, there are still several aspects of UDGs that remain intriguing.
Notably, UDGs on average have more globular clusters (GCs) than normal galaxies of similar stellar mass \citep{Dokkum16,Dokkum17,Lim18,Forbes20}.
There exists an empirical relation between GC abundance and dark-matter (DM) halo mass, valid over almost five dex of virial mass for normal galaxies \citep{SF09,Hudson14,Harris15,Harris17,BF20}. 
If this relation holds for UDGs, then a higher-than-average GC abundance implies that the host DM halo is overly massive for the stellar mass.
In contrast, some UDGs seem to be DM deficient, based on their GC or gas kinematics \citep{Dokkum18,Dokkum19,Danieli19,Guo20}, which poses a challenge to the standard picture where galaxy formation takes place in DM halos that dominate the mass budget. 
Hence, to understand the GC populations of UDGs takes center stage in understanding UDG formation in a cosmological context. 
Notably, the abundance, the spatial distribution, and the kinematics of GCs all contain information of the DM distribution of their hosting UDG. 

The galaxy NGC5846-UDG1 (UDG1 hereafter) serves as a remarkable example \citep{Forbes19,Muller20,Forbes21, Danieli22,Bar22}.
On the one hand, it hosts a surprisingly large population of GCs (of $\sim 50$ within two stellar effective radii) for its stellar mass of $\sim10^{8}\Msun$.
This translates to an overly massive DM halo of $\sim10^{11}\Msun$ assuming the \citeauthor{Harris17} relation \citep{Forbes21}. 
On the other hand, the GC population of UDG1 shows a strong radial mass segregation \citep{Bar22}, with more massive GCs lying closer to the center of the galaxy.
The mass segregation can be most naturally interpreted as a manifestation of orbital decay caused by dynamical friction (DF), because the strength of DF scales strongly with the perturber mass. 
And if DF causes the mass segregation, the halo mass should be much lower, because the timescale of orbital decay depends on the perturber-to-host mass ratio ($m/M$), such that it is shorter than the dynamical timescale of the host galaxy only if $m/M \ga 1/100$ \citep[e.g.][]{BK08}. 
That is, for a GC of mass $m \sim 10^6\Msun$, the host halo cannot be orders of magnitudes more massive than the stellar component ($\sim10^8\Msun$) in order to have sufficient orbital decay and mass segregation. 
Admittedly, this mass-ratio argument was originally made for satellite galaxies entering the host at orbital energies comparable to that of a circular orbit at the virial radius, so the GCs near the host center and thus with much lower orbital energy in principle allow for smaller mass ratios (and therefore larger host halo masses).
This, however, requires more detailed modeling that considers the locations of the GCs at birth and the density profile of the host. 
The strikingly different halo-mass estimates based on the aforementioned two perspectives highlights the importance of such models.

In this work, we first present a semi-analytical model of GC evolution in a composite host potential consisting of stellar and DM distributions. 
While generally applicable to any dwarf galaxy that exhibits a radial trend of its GC properties, here this model is applied to UDG1 as a proof of concept, showing that the observed mass segregation, together with the other information of the GCs available in the imaging data, can be used to constrain the DM halo.
As a major improvement over previous studies which also attribute the GC mass segregation to dynamical friction \citep{Bar22,Modak22}, a more physical model of the evolution of star clusters under the influence of tidal interactions with the host galaxy and the internal two-body relaxation is considerted.  
Tidal interactions and two-body relaxation drive mass loss and structural changes of the GCs, thus affecting the orbital evolution and the spatial distribution in a subtle but important way, as we will discuss below. 
When combined with parameter inference tools, this model enables using imaging data alone, without costly kinematics observations, to statistically constrain the DM distribution of the host galaxy. 
Additionally, the extreme limit of GC mass segregation is the complete orbital decay of massive GCs to the galaxy center, which is a viable way of forming dense nuclear star clusters as observed in nucleated low-surface-brightness galaxies \citep{Lim18,Greco18,SJ19,Iodice20,Marleau21}.  
Our method is therefore also potentially useful in this context.

\begin{figure*}
    \centering
    
    \tikzstyle{block} = [rectangle, draw, fill=blue!20,
    text width=16em, rounded corners, minimum height=4em]
    \tikzstyle{blockgray} = [rectangle, draw, fill=gray!20,
    text width=12em, rounded corners, minimum height=4em]
    \tikzstyle{narrowblock} = [rectangle, draw, fill=blue!20,
    text width=13em, rounded corners, minimum height=4em]
    \tikzstyle{blockgreen} = [rectangle, draw, fill=green!20,
    text width=12em, rounded corners, minimum height=4em]
    \tikzstyle{blockred} = [rectangle, draw, fill=red!20,
    text width=12em, rounded corners, minimum height=4em]
    \tikzstyle{blockorange} = [rectangle, draw, fill=orange!20,
    text width=12em, rounded corners, minimum height=4em]
    \tikzstyle{line} = [draw, thick, ->]
    
    \begin{tikzpicture} [node distance = 2cm]
    
        \node [blockred, text width=17em] (set) {
        \textbf{Input fixed parameters:}\\
        ICMF mass scales $m_{\rm min}$ \& $m_{\rm max}$ \\
        Evolution time: 10 Gyr (age of GCs)
        };
        
        \node [blockred, text width=18em, below of=set, node distance=2cm,] (free) {
        \textbf{Input free parameters:}\\
        Halo mass $M_{\rm h}$\\
        Halo concentration $c$\\
        Scale radius of intial GC distribution $r_0$};
        
        \node [block, right of=set, text width=30em, node distance=1cm,xshift=9cm,yshift=-1cm] (init) {
        \textbf{Initialization (Section~\ref{sec:assumptions}):}\\
        Mass: drawn from the ICMF $\frac{\rmd N}{\rmd m}\propto m^{-2} \exp(-\frac{\mmin}{m})\exp(-\frac{m}{\mmax})$   \\
        Size: $\lhalf = 2.55\kpc (m/10^4\Msun)^{0.24}$ \citep{BG21} \\
        Profile: EFF profile \citep{Elson87}\\
        Orbit: isotropic in both spatial distribution and velocity distribution, with distances drawn from a Hernquist profile of scale radius $r_0$ and velocities drawn from the Eddington's inversion method};
        
        \node [block, below of=init, node distance=4cm, xshift=-3cm, text width=15em] (orbit) {
        \textbf{Orbit evolution (Section~\ref{sec:DF}):}\\
        Evolve in a composite host potential consisting of a DM halo (NFW or Burkert profile) and a stellar density profile (Appendix \ref{sec:bgprofile}), with dynamical friction \citep{Chandrasekhar43,Petts15,Petts16}};
        
        \node [block, right of=orbit, node distance=2cm, xshift=3.5cm, text width=15em] (mass-size) {
        \textbf{Mass-size evolution (Section~\ref{sec:mass-size}):}\\
        Improved upon \citet{GR16}, with mass loss and structural relaxation in response to two-body relaxation and tidal interactions (following the tidal tracks of \citet{Penarrubia10} in tidally limited cases) };
        
        \node [block, below of=mass-size, node distance=3cm, xshift=-2cm, text width=16em] (record) {
        \textbf{Record the observables:}\\
        Present-day mass $m$\\
        Present-day half-mass radius $l_{\rm 1/2}$\\
        Present-day coordinates $\bm{r}$ \& $\bm{V}$};
        
        \node [blockorange, left of=record, node distance=6cm,] (MCMC) {
        \textbf{MCMC (Section~\ref{sec:MCMC}):}\\
        Likelihood based on the evolved GC mass, size, and spatial distributions};
        
        \node [blockgreen, left of=MCMC, node distance=5cm,] (output) {
        \textbf{Posterior (Section~\ref{sec:results}):}\\
        Halo mass $M_{\rm h}$\\
        Halo concentration $c$\\
        Scale radius $r_0$};
        
        \path [line] (free) -- (init);
        \path [line] (set) -- (init);
        \path [line] (init) -- (mass-size);
        \path [line] (init) -- (orbit);
        \path [line] (mass-size) -- (record);
        \path [line] (orbit) -- (record);
        \path [line] (record) -- (MCMC);
        \path [line] (MCMC) -- (free);
        \path [line] (MCMC) -- (output);

    \end{tikzpicture}
    \caption{Model workflow. The {\tt SatGen} \citep{Jiang21} semi-analytical framework for galaxy evolution provides the backbone of this model. The star-cluster-specific prescriptions are summarized here and detailed in \se{model}. }
    \label{fig:schematic}
\end{figure*}
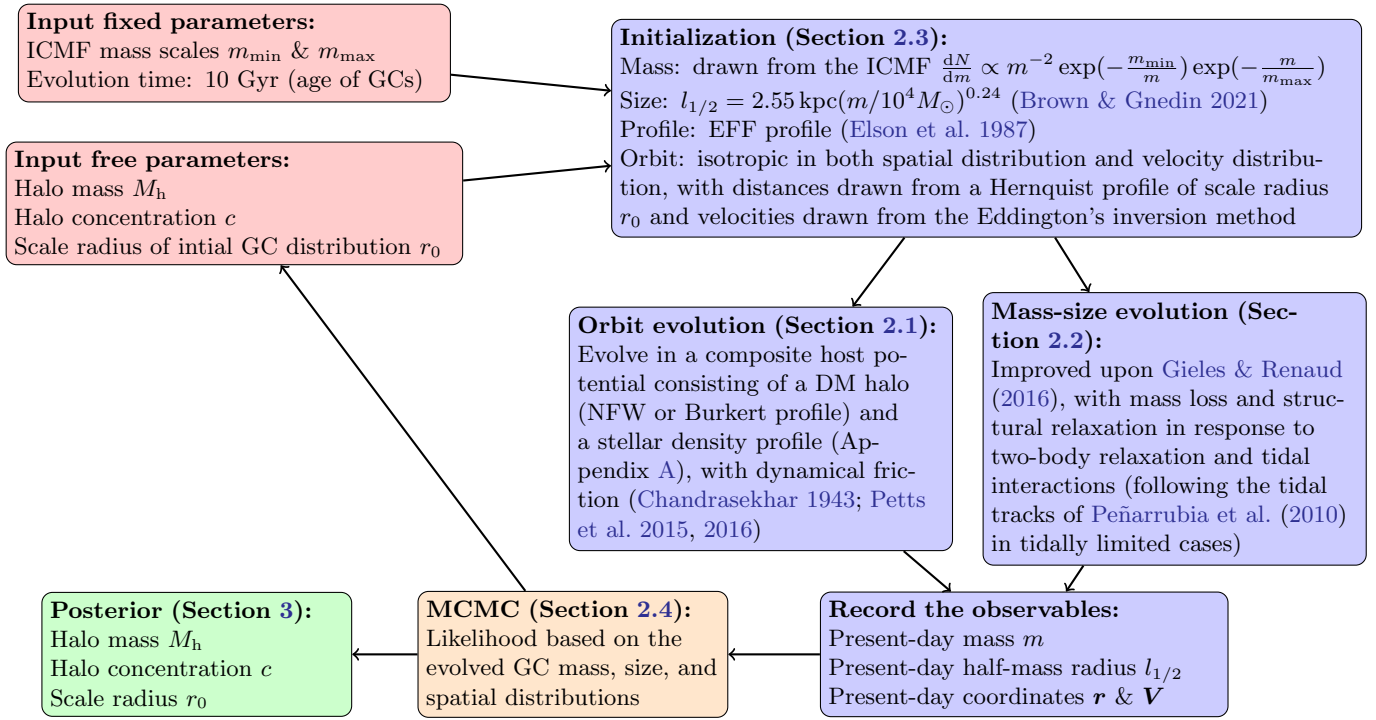

This work is organized as follows. 
In Section \ref{sec:model}, we introduce our model of GC evolution, and present the workflow of forward modeling the GC population and inferring the host DM profile. 
For readers who wish to skip the technical details, \Fig{schematic} presents a schematic flowchart that summarizes all the components of our framework, and serves as a self-contained starting place before reading the result sections. 
In Section \ref{sec:results}, we use the observed GC statistics of UDG1 to constrain the model parameters, including the DM halo mass and concentration, as well as the characteristic spatial scale of the initial distribution of the GCs which may shed light upon the origin of the GCs. 
In Section \ref{sec:discussion}, we compare the model predictions and kinematics observations, discuss the key distinction between a cuspy halo profile versus a cored profile regarding star-cluster statistics, compare our model to simplistic models that ignore the physics of GC mass and structural evolution, and also comment on potential future developments of this methodology. 
We draw our conclusions in Section \ref{sec:conclusion}.

Throughout, we define the virial radius of the hosting DM halo as the radius within which the average density is $\Delta=200$ times the critical density for closure, and adopt a flat cosmology with the present-day matter density $\Omega_{\rm m}=0.3$, baryonic density $\Omega_{\rm b}=0.0465$, dark energy density $\Omega_\Lambda=0.7$, a power spectrum normalization $\sigma_8=0.8$, a power-law spectral index of $n_s=1$, and a Hubble parameter of $h=0.7$. We use $r$, $R$, and $l$ to indicate the three-dimensional galactocentric radius, the projected galactocentric radius, and star-cluster-centric radius, respectively; and denote the mass of a star cluster and that of the host galaxy by $m$ and $M$, respectively.
\\

\section{Methodology} \label{sec:model}

In this section, we first introduce a dynamical model that describes the evolution of GCs in a composite host potential consisting of dark matter and stars. 
This model considers the orbital evolution of GCs under the influence of dynamical friction (DF), allowing for the dependence of the strength of DF on the local density profile of the host potential following the recipe of \citet{Petts15}.
The GCs evolve in mass and structure in response to the internal two-body relaxation and the varying tidal effects along the orbits.
We then lay out the model assumptions, including the initial star-cluster mass function, the initial structure of young star clusters, the initial spatial distribution of the star clusters, and the assumptions about the host potential. 
With these assumptions, the model self-consistently evolves a population of GCs, predicting their evolved masses, sizes, and spatial distribution, which are then compared to those of the observed GC population. 
Finally, we combine the model with a Markov Chain Monte Carlo (MCMC) inference tool, to derive constraints on the DM halo of the target galaxy. 
The workflow of our model is summarized in \Fig{schematic}.

\subsection{Orbit Evolution}\label{sec:DF}

To follow the orbit of a GC, we solve the equation of motion,
\begin{equation}
    \bm{\Ddot{r}}=-\nabla\Phi+\bm{a}_{\text{DF}},
\end{equation} 
where $\bm{r}$ is the position vector, $\Phi$ is the gravitational potential, and $\bm{a}_{\text{DF}}$ is the acceleration due to DF, given by \citep{Chandrasekhar43}
\be\label{eq:DF}
   \bm{a}_{\text{DF}}=-4\pi G^2m\sum_i\ln\Lambda_i\rho_i(\bm{r})F(<V_i)\frac{\bm{V}_i}{V_i^3}.
\ee
Here, the summation is over different components of the host potential ($i$ = DM, stars), $m$ is the mass of the GC, $\ln\Lambda_i$ is the Coulomb logarithm, $\bm{V}_i$ is the relative velocity of the GC with respect to component $i$, and $F(<V_i)$ is fraction of particles that contribute to DF, which, under the assumption of a Maxwellian distribution, is given by
$F(<V_i)=\text{erf}\left(X_i\right)-(2X_i/\sqrt{\pi})e^{-X_i^2}$,
where $X_i=V_i/(\sqrt{2}\sigma_i)$, with $\sigma_i$ the one-dimensional velocity dispersion of component $i$ at position $\bm{r}$.

In the idealized \citeauthor{Chandrasekhar43} picture of dynamical friction, the perturber travels across an infinite homogeneous isotropic medium, and $\Lambda$ is defined as $b_{\rm max}/b_{\rm min}$, with $b_{\rm max}$ and $b_{\rm min}$ the maximum and minimum impact parameters, respectively. 
For perturbers orbiting a galaxy, which is not a uniform medium, the \citeauthor{Chandrasekhar43} DF treatment is used as an approximation, where $b_{\rm max}$ is of the order of the characteristic size of the host system, and $b_{\rm min}$ is the larger of the impact parameter for a 90-degree deflection and the size of the perturber \citep{BT08}. 
In semi-analytical models of satellite-galaxy evolution, it is a common practice to simply  assume $\lnL \sim \ln(M/m)$, where $M/m$ is the mass ratio of the host and the satellite, as the virial radius of a gravitationally bound structure scales with the virial mass \citep[see e.g.,][and the references therein]{Gan10}. 
Even constant Coulomb logarithms of $\lnL\sim 3$ are widely adopted, as major and minor mergers ($M/m\la 10$) contributes to the bulk of the surviving satellite galaxies. 
Hence, for the purpose of studying satellite galaxies, where typically the focus is not on the orbital evolution of individual perturbers but on the overall satellite statistics, the simplistic forms of Coulomb logarithm such as $\lnL\sim\ln(M/m)$ and $\sim3$ are reasonable \citep{Green21}. 
However, for our purpose here, i.e., to use the GC mass segregation to constrain the dynamical mass distribution, the details of individual orbits are important and thus the simplistic Coulomb logarithms for satellite galaxies may be problematic. 
For example, $\lnL \sim \ln(M/m)$ would be very high for GCs and the orbital decay would be unrealistically strong. 

Hence, following the more detailed treatment of \citet{Petts15}, we choose $\bmax$ to be
\be\label{eq:bmax}
\bmax(r) = \min\left\{\frac{r}{\gamma(r)},r\right\}
\ee
where $\gamma(r)\equiv -\rmd\ln\rho/\rmd\ln r$ is the local logarithmic density slope of the host potential, and choose $\bmin$ as
\be\label{eq:bmin}
\bmin = \max\left\{\lhalf,\frac{Gm}{V^2}\right\},
\ee
where $\lhalf$ is the half-mass radius of the GC. \footnote{In practice, we do not distinguish the half-mass radius from the effective radius (i.e., 2D half-light radius), which is provided by the observational data for the GCs in UDG1 and the size-mass relation of young star clusters \citep[][as will be discussed in \se{assumptions}]{BG21}.}  
As such, $\bmax$ is a length scale over which the density is approximately constant \citep{Just11}.
To deal with the cases of $\bmax\sim\bmin$, which can happen when a GC approaches the center of the host, we use the original \citeauthor{Chandrasekhar43} result for the Coulomb logarithm $0.5\ln(\Lambda^2+1)$ in place of the $\lnL$ in \eq{DF}.
These choices empirically account for the core-stalling effect \citep{Goerdt06,Read06,Inoue09,KS18}, the phenomenon that the DF acceleration decreases and the orbital decay stalls when the perturber approaches a flat density core -- because in the density core, $\gamma(r) \sim 0$, $\bmax\sim\bmin\sim r$ and $\ln(\Lambda^2+1)\sim0$.

\begin{figure}	
\includegraphics[width=\columnwidth]{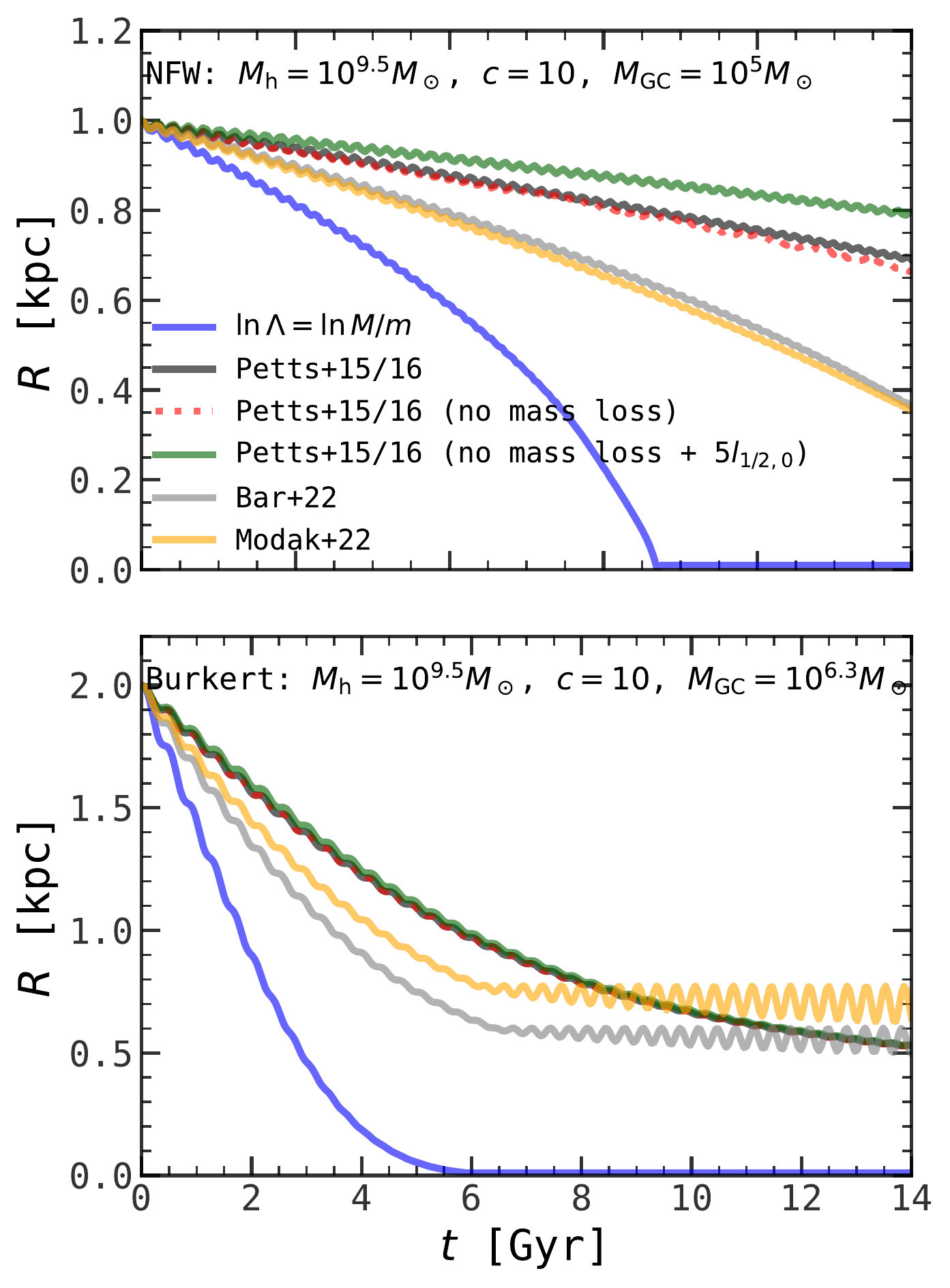}
    \caption{
    Illustration of the effects of different prescriptions of the Coulomb logarithm in the \citeauthor{Chandrasekhar43} DF formula on the orbit evolution of a star cluster in a cuspy NFW host (upper) and in a cored Burkert host (lower), respectively. The parameters of the host halos and star clusters are chosen to better reveal the differences between different models, as indicated. The GCs are released at $t=0$ on almost circular orbits.
    The black line stands for the result using our fiducial model of GC evolution (\se{mass-size}) and the \citeauthor{Petts15} Coulomb logarithm (\se{DF}). 
    The other lines represent the results from varying certain aspect, using, e.g., the same fiducial $\Lambda$ but with no mass loss (red), the same $\Lambda$ but with no mass loss and with the GC size boosted by 5x (green), the fiducial evolution model but with a $\Lambda$ widely assumed for satellite-galaxy evolution (blue).
    Note that, first, the GC mass-size evolution affects orbital evolution via the DF treatment; second, the \citeauthor{Petts15} Coulomb logarithm gives much weaker orbital decay than the simplistic $\lnL=\ln M/m$;
    third, the \citeauthor{Bar22} and \citeauthor{Modak22} recipes, which are simplified versions of \citeauthor{Petts15} that ignore the GC size in $\bmin$, result in stronger DF than the full treatment.}
    \label{fig:DF}
\end{figure}

\Fig{DF} compares the orbital evolution for different prescriptions of the Coulomb logarithm as well as for different GC mass evolution models, in an NFW host halo and in a Burkert halo, respectively. 
Focusing on the comparison of the blue and black lines in either host profile, we can see that the simplistic Coulomb logarithm of $\ln(M/m)$ yields significantly stronger orbital decay than the more accurate \citeauthor{Petts15} treatment, which has been well calibrated against numerical simulations. 

We also note that in two previous studies that are highly relevant, \citet{Bar22} and \citet{Modak22}, the Coulomb logarithms were chosen to be simplified variants of what we use here. 
The key difference is that their models do not follow the size evolution of the GCs, so their $\bmin$ are effectively $\sim Gm/V^2$, which is usually a factor of a few smaller than $\lhalf$, making the Coulomb logarithm larger and DF stronger than the full treatment, in the NFW case as illustrated in the upper panel of \Fig{DF}. 
In the Burkert case, \citeauthor{Bar22} model core stalling by setting Coulomb logarithm zero when GCs get within 0.3 times the half-mass radius of the galaxy.  
We will elaborate on the comparisons of our model with these models in \se{comparison}.

\subsection{A unified model of mass-size evolution}\label{sec:mass-size}

GCs are compact objects that are more resilient to the environmental processes than more diffuse substructures of a galaxy such as DM substructures and gas clouds.
This is largely why in several previous studies of GC mass segregation, the mass evolution of the GCs was treated simplistically and the structural evolution of the GCs were completely ignored \citep{Dutta19,Bar22,Modak22}.
However, when dealing with the long-term evolution over the age of the clusters ($\sim10\Gyr$), the tidal interactions between the GCs and the host galaxy, especially near the center of the host, can result in non-trivial mass and structural change. 
In the meantime, GCs are internally collisional, and thus lose mass and expand due to the evaporation of stars. 
This is relevant even in low-density environments. 
The combination of the external tidal effects and the internal two-body relaxation may result in non-linear mass loss and structural change, which, in turn, affects the orbital evolution, since the DF acceleration depends on the mass and structure of the perturber, as discussed in \se{DF}. 
Additionally, the imaging data of the GCs in nearby low-surface-brightness galaxies can be high-resolution enough to provide information about the internal structure of the GCs \citep{Dokkum18}. 
This potentially also contains valuable information of the dynamics besides the mere mass and spatial distribution. 
For all these reasons, we model the mass-size evolution of the GCs. 

Our model of GC mass-size evolution adopts a similar formalism as that of \citet[][GR16]{GR16} but is different in two important aspects. 
First, GR16 focused on the evolution of newly born star clusters younger than 100 Myr in the vicinity of their birth places, and therefore the dominant tidal effect is the repeated impulsive encounters with giant molecular clouds in the clumpy inter stellar medium surrounding the clusters. 
Here, we trace the long-term evolution of star clusters over cosmological timescales, and
therefore we focus more on the tidal interactions with the background potential. 
Second, because of the short-term nature, the GR16 model assumes that two-body relaxation causes no mass loss, whereas here we cannot ignore the mass loss from two-body relaxation over the age of the GCs.
We assume that star clusters follow the \citet[][EFF]{Elson87} density profile with an outer density slope of $-4$, and the shape is fixed across evolution such that the density-profile evolution is manifested only by the mass and size evolution laid out below. 

We start by differentiating the binding energy $E\propto - G m^{5/3}\rhohalf^{1/3}$ of a GC, and express the derivative in terms of the mass, $m$, and the average density within the half-mass radius, $\rhohalf$:
\be
\frac{\dd E}{E} = \frac{5}{3} \frac{\dd m}{m} +  \frac{1}{3} \frac{\dd \rhohalf}{\rhohalf}.
\ee
Both tidal interactions and two-body relaxation contribute to the energy increase $\dd E$ and the mass loss $\dd m$, so we distinguish their contributions by denoting $\dd E$ and $\dd m$ in two terms with subscripts ``t'' and ``r'', respectively, 
\be\label{eq:DifferentiatingEnergy}
\frac{\dd \Et+\dd \Er}{E} = \frac{5}{3} \frac{\dd \mt + \dd \mr}{m} +  \frac{1}{3} \frac{\dd\rhohalf}{\rhohalf}.
\ee
Following GR16, we introduce a parameter $\ft$ to relate the mass loss to the tidal heating from the interactions with the host potential:
\be\label{eq:Defineft}
\frac{\dd \mt}{m} = \ft \frac{\dd \Et}{E}.
\ee
Similarly, we define an $\fr$ parameter that relates mass loss to the internal heating due to two-body relaxation:
\be\label{eq:Definefr}
\frac{\dd \mr}{m} = \fr \frac{\dd \Er}{E}.
\ee
The values of $\ft$ and $\fr$ can be estimated following analytical arguments or empirical numerical results, as will be elaborated shortly. 
It is easier to model the mass losses than to model the energy changes, so we proceed by eliminating the energy terms in \eq{DifferentiatingEnergy} using the definitions of $\ft$ and $\fr$. 

The mass loss from tidal stripping is computed as
\be\label{eq:TidalStrippingMassLoss}
\frac{\dd \mt}{m} = -\alpha \xit \frac{\dd t}{\taudyn},
\ee
where $\xit\equiv [ m-m(\lt)]/m $ is the fraction of mass outside the tidal radius, with $m(\lt)$ the mass within the instantaneous tidal radius, $\lt$; $\taudyn$ is the dynamical time of the host potential at the GC's instantaneous location, given by
\be\label{eq:taudyn}
\taudyn = \sqrt{\frac{3\pi}{16G\bar{\rho}(r)}},
\ee
with $\bar{\rho}(r)$ the average density of the host system within radius $r$; and $\alpha\approx0.55$ is an empirical coefficient, calibrated with $N$-body simulations \citep{Green21}.
The tidal radius is given by \citep{King62}
\be\label{eq:lt}
\lt = r\left[\frac{m(\lt)/M(r)}{2-\dd\ln M/\dd\ln r
+ \Vt^2/\Vc^2(r)}\right]^{1/3},
\ee
where $M(r)$ is the host mass within radius $r$, $\Vt=|\hat{r}\times \bm{V}|$ is the instantaneous tangential velocity, and $\Vc(r)$ is the circular velocity.

Similarly, the evaporation caused by two-body relaxation can be expressed as
\be\label{eq:RelaxationMassLoss}
\frac{\dd m_{\rm r}}{m} = - \xi_{\rm e} \frac{\dd t}{\tau_{\rm r}},
\ee
where $\xi_{\rm e}\equiv [ m-m(<v_{\rm esc})]/m$ is the fraction of stars in the tail of the velocity distribution that is larger than the escape velocity, which, for an isolated relaxed GC and thus a Maxwellian velocity distribution, is a constant $\xi_{\rm e}\approx0.0074$; and $\tau_{\rm r}$ is a relaxation timescale, given by \citep{Spitzer87,GR16}
\be\label{eq:RelaxationTime}
\tau_{\rm r} \approx 0.142 \text{Gyr} \left( \frac{m}{10^4 M_\odot}\right) \left( \frac{\rhohalf}{10^{11}M_\odot\text{kpc}^{-3}} \right)^{-1/2}.
\ee
This is the timescale of refilling the high-speed tail of the velocity distribution.

Combining \eqs{DifferentiatingEnergy}-(\ref{eq:RelaxationMassLoss}), we obtain a unified model for GC structural evolution
\be\label{eq:UnifiedModel}
\frac{\dd \rhohalf}{\rhohalf} = \left[ \alpha \left(5-\frac{3}{f_{\rm t}}\right)\frac{\xi_{\rm t}}{\tau_{\rm dyn}} +  \left(5-\frac{3}{f_{\rm r}}\right) \frac{\xi_{\rm e}}{\tau_{\rm r}} \right] \dd t.  
\ee
The parameters on the right-hand side of \eq{UnifiedModel} all have analytical estimates or empirical values based on numerical simulations. 

GR16 adopted $\ft=3$ and $\fr=0$ for the short-term evolution of young star clusters in clumpy interstellar medium; here we estimate $\ft$ and $\fr$ in the context of the long-term evolution of GCs in a gas-less host.
To estimate $\ft$, we consider the limit of negligible two-body relaxation, where the GC evolution can be approximated by the tidal evolution of self-gravitating collisionless systems, which has been extensively studied in the context of DM subhalos \citep[e.g.,][]{Penarrubia10, BD22}.
Notably, \cite{Penarrubia10} calibrated the tidal evolutionary tracks for DM subhalos using $N$-body simulations in terms of the maximum circular velocity $v_{\rm max}$ and the corresponding radius $l_{\rm max}$ as functions of the bound mass fraction $x=m(t)/m(0)$ and the inner logarithmic density slope $s=-\dd \ln\rho/\dd\ln l |_{l\to0}$.
Turning off two-body relaxation by setting the second term of \eq{UnifiedModel} to zero, i.e., 
\be\label{eq:ModelWithoutRelaxation}
\frac{\dd \rhohalf}{\rhohalf} =  \alpha \left(5-\frac{3}{f_{\rm t}}\right) \xi_{\rm t} \frac{\dd t}{\tau_{\rm dyn}},
\ee
we can therefore find $f_{\rm t}$ by matching the structural evolution according to \eq{ModelWithoutRelaxation} in terms of $v_{\rm max}$ and $l_{\rm max}$ to the tidal track of \citeauthor{Penarrubia10} for the case of $s=0$, since young star clusters are generally well described by the \citet{Elson87} profile and have flat density cores (see \se{assumptions}).
We find that $f_{\rm t}$ is of order unity and mildly decreases with the bound mass fraction:
\be\label{eq:ft}
f_{\rm t} = 0.77 x^{0.19}, \, x= m/m(0).
\ee

To estimate $\fr$, we follow \cite{Gieles11} and the seminal work of \cite{Henon65} to express the energy change of an isolated GC due to two-body relaxation as
\be\label{eq:RelaxationEnergyChange}
\frac{\dd E_{\rm r}}{E} = - \zeta \frac{\dd t}{\tau_{\rm r}},
\ee
where $\zeta\approx0.0926$, assuming equal stellar masses of $0.5 M_\odot$ and a Coulomb logarithm of $10$ within the star cluster.\footnote{For any realistic stellar mass spectrum, the $\zeta$ parameter is larger, up to $\sim0.5$ as discussed in GR16.} 
Comparing \eqs{Definefr}, (\ref{eq:RelaxationMassLoss}), and (\ref{eq:RelaxationEnergyChange}), we obtain
\be\label{eq:fr}
f_{\rm r} = \xi_{\rm e}/\zeta \approx 0.08.
\ee

\begin{figure}	
\includegraphics[width=1.05\columnwidth]{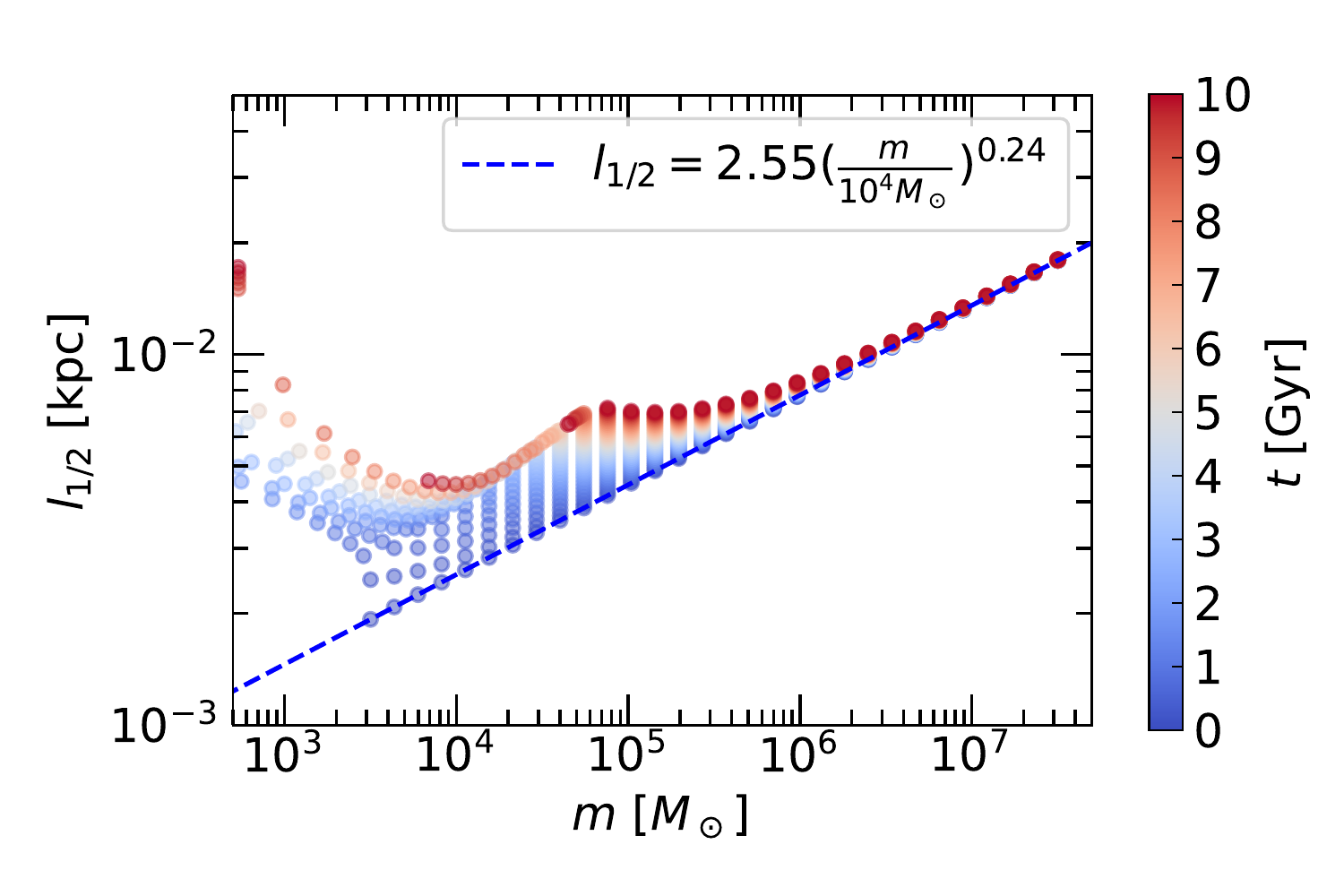}
    \caption{
    Illustration of the mass-size evolution of star clusters over 10 Gyr according to the model presented in \se{mass-size}. The star clusters are initialized with masses uniform in $\log m$, sizes following the observed size-mass relation of young star clusters \citep[][blue dashed line]{BG21}, and circular orbits of $r=5$ kpc in a host potential consisting of an NFW halo with $\Mh=10^{12}\Msun$ and $c =10$ and a UDG1-like stellar profile.
    The evolution is mass-dependent, with massive clusters almost intact and low-mass clusters expanding first and then quickly getting tidally truncated and disrupted.}
    \label{fig:size-mass}
\end{figure}

In summary, for each timestep along the orbit, we evolve the mass of a GC using \eqs{TidalStrippingMassLoss} and (\ref{eq:RelaxationMassLoss}), and update the structure of the GC using \eq{UnifiedModel}, with the parameters $\alpha=0.55$, $\fr=0.08$, and $\ft$ given by \eq{ft}.
The initial mass and structure of a GC is chosen according to the assumptions that will be given in \se{assumptions}.  

\Fig{size-mass} illustrates the behavior of GCs in the size-mass plane over 10 Gyr according to this model. 
The GCs are initialized with masses uniformly distributed in logarithmic mass and with sizes according to the median observed size-mass relation of young star clusters \citep[][as will be discussed in \se{assumptions}]{BG21}.
Clearly, the evolution depends on the initial mass.
For the most massive GCs ($m \ga 10^6\Msun$), the size and mass barely evolve. 
For intermediately massive GCs ($m \approx 10^{5-6}\Msun$), the main effect is expansion due to two-body relaxation, while the mass evolution is marginal.  
Basically, within 10 Gyr, the expansion barely makes their mass distribution extend to the tidal radius, so there is almost no tidal truncation.  
For lower-mass clusters ($m\la 10^{4.5}\Msun$), tidal truncation quickly takes effect as they expand, so they start to lose mass quickly and even dissolve. 
These mass-dependent behaviors work together to shape the evolved GC mass function in  qualitatively the correct direction to that observed, peaking at $m\sim10^5 M_\odot$. 
The evolved GC mass distribution would be insensitive to the low-mass end of the initial cluster mass function.
With low mass GCs stripped and dissolved, intermediate mass GCs experiencing weak DF, and massive GCs largely intact and thus always experiencing the strongest DF, mass segregation would naturally arise. 

\subsection{Model assumptions and GC initialization}\label{sec:assumptions}

We emphasize that the scenario that this study focuses on is that the observed GC mass-segregation signal is driven by DF and that the strength of it can be used to constrain the DM distribution the host galaxy. 
Hence, the assumptions are chosen to keep the setup simple and to serve the purpose of testing the constraining power of GC mass segregation on DM halo properties. 
We assume that the initial cluster mass function (ICMF) follows a power-law with exponential truncations at both the high-mass and low-mass ends \citep{Trujillo-Gomez19}:
\be\label{eq:ICMF}
\frac{\text{d}N}{\text{d}m}\propto m^\beta\text{exp}\left(-\frac{m_{\rm min}}{m}\right)\text{exp}\left(-\frac{m}{m_{\rm max}}\right),
\ee
where $\beta=-2$ reflects hierarchical molecular cloud formation, and $m_{\min}$ and $m_{\rm max}$ are the lower and upper characteristic scales. 
We keep $\mmin=10^{5.5} M_\odot$ and $\mmax=10^{8} M_\odot$ fixed for simplicity, after having verified that the results are not sensitive to the detailed values as long as they allow for the existence of GCs covering the mass range of $\sim 10^{4.5}$ to $10^6\Msun$ 
\footnote{We emphasize that $\mmin$ should not be regarded as the minimum initial mass: because of the functional form of \eq{ICMF} and the power-law slope of $\beta=-2$, with $\mmin=10^{5.5}\Msun$, the minimum initial star-cluster mass can reach $\sim10^4\Msun$. Similarly, $\mmax$ is not the maximum GC mass.
}.
This is partly due to the mass-dependent evolution as shown in \Fig{size-mass}, that the low-mass clusters will dissolve in the end. 

We describe the density profile of a star cluster by an \citet[][EFF]{Elson87} functional form,
\be\label{eq:EFF}
\rho(l) = \frac{\rho_0}{(1+l^2/a^2)^\eta},
\ee
where $\rho_0 = \Gamma(\eta)m/[\pi^{3/2}\Gamma(\eta-1)a^3]$ is the central density,
with $\Gamma(x)$ the Gamma function, $m$ the mass of the cluster, $-2\eta$ the outer logarithmic density slope, and $a$ a scale radius linked to the half-mass radius by
\be\label{eq:EFFHalfMassRadius}
a =\lhalf / (2^{\frac{2}{2\eta-3}}-1)^{1/2}.
\ee
We adopt $\eta=2$ such that the outer density slope is $-4$, motivated by observations of the light profile of young star clusters \citep{Ryon15}, and assume that the slope is fixed across evolution such that the density-profile evolution is manifested only by the mass and size evolution as in \se{mass-size}. 
More analytical properties of the EFF profile are presented in Appendix \ref{sec: EFFprofile}. 

With the initial mass drawn from the ICMF, we determine the initial size by sampling from a log-normal distribution based on the observed size-mass relation for young star clusters in the Legacy Extragalactic UV Survey \citep{BG21}: the median half-mass radius is given by
\be\label{eq:size-mass}
\lhalf=2.55\kpc\left(\frac{m}{10^4 M_\odot}\right)^{0.24},
\ee
and the $1\sigma$ scatter at fixed mass is approximately 0.25 dex. 

We assume that the GC progenitors were all born 10 Gyr ago \citep{Muller20,Bar22} and that they were isotropically distributed following a \citet{Hernquist90} profile, $\rho(r)\propto 1/[r(r+r_0)^3]$, with no mass segregation at birth.
We briefly explore the possibility of an initial mass segregation in \app{init_segre}. 
The initial spatial scale, $r_0$, of the GC-progenitor spatial distribution is a free parameter to be constrained.  
We note that the GC-GC merger rate is only $\sim 0.03 \Gyr^{-1}$ per GC assuming $\sim100$ GCs in a dwarf halo of $10^{10}\Msun$, using the merger criterion in \cite{Dutta20}.
We have also numerically verified, using the GC-merger prescription of \citet{Modak22}, that for such a dwarf halo, GC mergers almost all occur at $r\la0.1\kpc$. 
Hence, to facilitate the MCMC inference, we practically ignore GC-GC encounters and mergers when focusing on the radial mass segregation signal at $r > 0.1\kpc$.  
For the GCs which have lost most of their orbital angular momenta and settle to $r<0.1\kpc$ before getting dissolved, we 
treat them as merging to form a nuclear star cluster (see \se{nucleated}). 

We treat the host system as a combination of a smooth stellar-mass distribution and a DM halo, both of which remain static during the GC evolution. 
For the stellar profile, to facilitate orbit integration, we fit a density profile with simple analytical expressions of the gravitational potential to the observed stellar density profile \citep{Bar22}, given by
\be
\label{eq:StellarDensityProfile}
\rho(r)=\frac{\rho_{0,\star}}{\left(1+x\right)\left(1+x^3\right)}
\ee
where $x=r/r_{\rm s}$, and $\rho_{0,\star} =27\Ms/\left[4\pi(9+2\sqrt{3}\pi)r_{\rm s}^3\right]$ with $\Ms = 10^{8.3} \Msun$ and $r_{\rm s}=2$ kpc (see \app{UDG1} for more details). 
For the DM halo, we consider representative functional forms for cuspy and cored profiles, respecitvely --  namely, the NFW \citep{Navarro97} profile,
\be
\rho(r) = \frac{\rho_0}{x\left(1+x\right)^2},
\ee
where $x=c r/\rv$, and $\rho_0 = c^3\Delta \rhoc/[3f(c)]$ with $ f(x)=\ln(1+x) - x/(1+x)$;
and the \citet{Burkert95} profile, 
\be\label{eq:BurkertDensity}
\rho(r) = \frac{\rho_0}{(1+x)\left(1+x^2\right)},
\ee
where $x=cr/\rv$, and $\rho_0 = \Mh/\left[2\pi \rv^3 g(c) c^3\right]$ with $g(x)=0.5\ln(1+x^2)+\ln(1+x) - \arctan x$.
It is not obvious whether a core or a cusp is more advantageous for producing the GC mass segregation: for a cored profile, GCs would pile up where the density slope turns flat due to the core-stalling effect, so that the massive GCs that sink to the core radius and the lower-mass GCs that were initially at the core radius are mixed; for a cuspy profile, DF could be so strong that massive GCs sink completely to the center, but leaving the outer GCs not very different in mass. 
It is therefore interesting to explore which case produces mass segregation more easily and what other differences they may cause. 

We initialize the velocities of the star clusters by sampling the velocity distribution function $\mathcal{P}(V|r)$ of a statistically steady-state system in absence of dynamical friction.  
Specifically, the ergodic energy distribution function is calculated from the Eddington's inversion method \citep{BT08}, 
\be\label{eq:Eddington}
f(\mathcal E)=\frac{1}{\sqrt{8} \pi^2}\left[\frac{1}{\sqrt{\mathcal E}}\left(\frac{\dd \rho}{\dd \Psi}\right)_{\Psi=0} +\int_0^{\mathcal E}\frac{\dd \Psi}{\sqrt{\mathcal E-\Psi}}\frac{\dd^2\rho}{\dd\Psi^2} \right],
\ee
where $\Psi(r)=-\Phi(r)$, with $\Phi$ and $\rho$ the gravitational potential and density profile of the DM halo. 
Then the conditional distribution of velocities at each radius $r$ is given by
\be\label{eq:vdistribution}
\mathcal{P}(V|r)=4\pi V^2\frac{f(\Psi(r)-V^2/2)}{\rho(r)}.
\ee
We draw the speeds from $\mathcal{P}(V|r)$ and assign the directions of the velocity vectors such that they are isotropic in space \citep[e.g., as detailed in][]{Jiang21}.

\subsection{Parameter Inference}\label{sec:MCMC}

We use the affine invariant MCMC ensemble sampler, {\tt emcee} \citep{Foreman-Mackey13}, to constrain the properties of the host DM halo (i.e. the halo mass $\Mh$ and the halo concentration parameter $c$) and the initial scale length of the GC distribution $r_0$. 
The observational data that provide the constraints involves the present-day masses of the GCs ($m$), the half-mass radii ($\lhalf$), and the projected distances to the galaxy center ($R$), from \citet{Danieli22}. 
\footnote{Note that the half-mass radii and masses were not published in their original work. We have used the full-width-at-half-maximum (FWHM) sizes to estimate the half-mass radii, as $\lhalf=\text{FWHM}/2$. We obtain the GC masses using $m=\gamma L_{V}$, where $\gamma$ is the mass-to-light ratio, set to be 1.6 \citep{Muller20}, and $L_{V}$ is V-band luminosity. We only include the GC candidates with $M_{V}<25$ mag, which gives us a sample of 34 GCs.}
With the primary focus being the radial segregation in mass, we adopt three logarithmic mass bins (as indicated in \Fig{NFWrealization}), and use the median quantities $\langle \log m \rangle_i$, $\langle \lhalf \rangle_i$, and $\langle \log R\rangle_i$ for constructing the likelihood.
We parameterize the radial mass segregation using two set of quantities: the slopes $\gamma_{ij}\equiv (\langle \log m\rangle_j - \langle \log m\rangle_i) / (\langle \log R \rangle_j - \langle \log R \rangle_i)$, and the number of GCs at each bin relative to the total number of the surviving GCs, $f_i$.  
These two quantities measure the steepness of the radial mass segregation and sample the evolved GC mass function, respectively. 
Overall, we consider a logarithmic likelihood given by
\be\label{likelihood}
    \ln(p)=-\frac{1}{2}\sum_k w_k \frac{\left(y_{k,\text{data}}-y_{k,\text{model}}\right)^2}{y_{k,\text{data}}^2},
\ee
where $y_{k,\text{data}}$ and $y_{k,\text{model}}$ refer to the observed values and model predictions, respectively, and  $y_k$ represents one of the following quantities, $\{\langle\log m\rangle_i\}$, $\{\langle\log R\rangle_i\}$ $\{\langle \lhalf\rangle_i\}$, $\{\langle\gamma\rangle_{ij}\}_{j>i}$, and $\{f_i\}$, with $i,j=1,2,3$ the mass-bin indices, and $w_k$ the weight for the $k$th quantity.
We adopt uniform weighting ($w_k=1$), which essentially gives the mass-segregation signal an emphasis because there are three $\gamma_{ij}$ slopes that measure it. 
We adopt uniform priors for $\log\Mh$, $\log c$, and the initial spatial scale $r_0$, within ranges that are chosen according to the galaxy of interest (see \se{results} for example). 

To speed up the MCMC inference, instead of evolving the GC populations on the fly for each iteration, we pre-compute the model predictions $y_{k,\text{model}}$ on a mesh grid spanned by the parameters of interest. 
During the MCMC random walk, $y_{k,\text{model}}$ is evaluated by linear interpolation. 
Examples of the pre-computed models can be seen in Appendix \ref{app:interpolation}.
Note that we opt for not including the total number of GCs as a quantity of interest in our model. 
This allows us to focus more efficiently on the correlations and on the moments of the observables.  
Hence, when pre-computing the models, we adopt arbitrarily large initial number of GCs to ensure smooth interpolations.
Note that ignoring GC mergers is inevitable in this approach, since merger rate depends on the total number.
That said, when presenting the model realizations corresponding to the posterior models, we adopt an initial GC number that leads to a surviving GC abundance comparable to what is observed. 


\section{The dark-matter halo of NGC5846-UDG1}\label{sec:results}

As a proof-of-concept, we apply the aforementioned method to study the halo of UDG1 and its GC population. 
We assume uniform priors of $\log(\Mh/\Msun)\in[8,10.5]$, $\log c\in[\log2,\log30]$, and $r_0\in[1,5]\kpc$, and choose $\Mmin=10^{5.5}\Msun$ and $\Mmax=10^8\Msun$ for the ICMF, and evolve  GCs for 10 \Gyr, after verifying that the results are not sensitive to slight variations of these values. 
We use 64 random walkers, and show results of 20000 iterations after 1000 burn-in timesteps.
Below, we first present the posterior distributions of the  parameters, then compare model realizations with the best-fit  parameters with the data, for the two halo-profile scenarios respectively, and finally discuss the results in the context of scaling relations of galaxy-halo connection. 

\subsection{NFW halo}\label{sec:NFWresult}

\begin{figure}	
\centering
\includegraphics[width=\columnwidth]{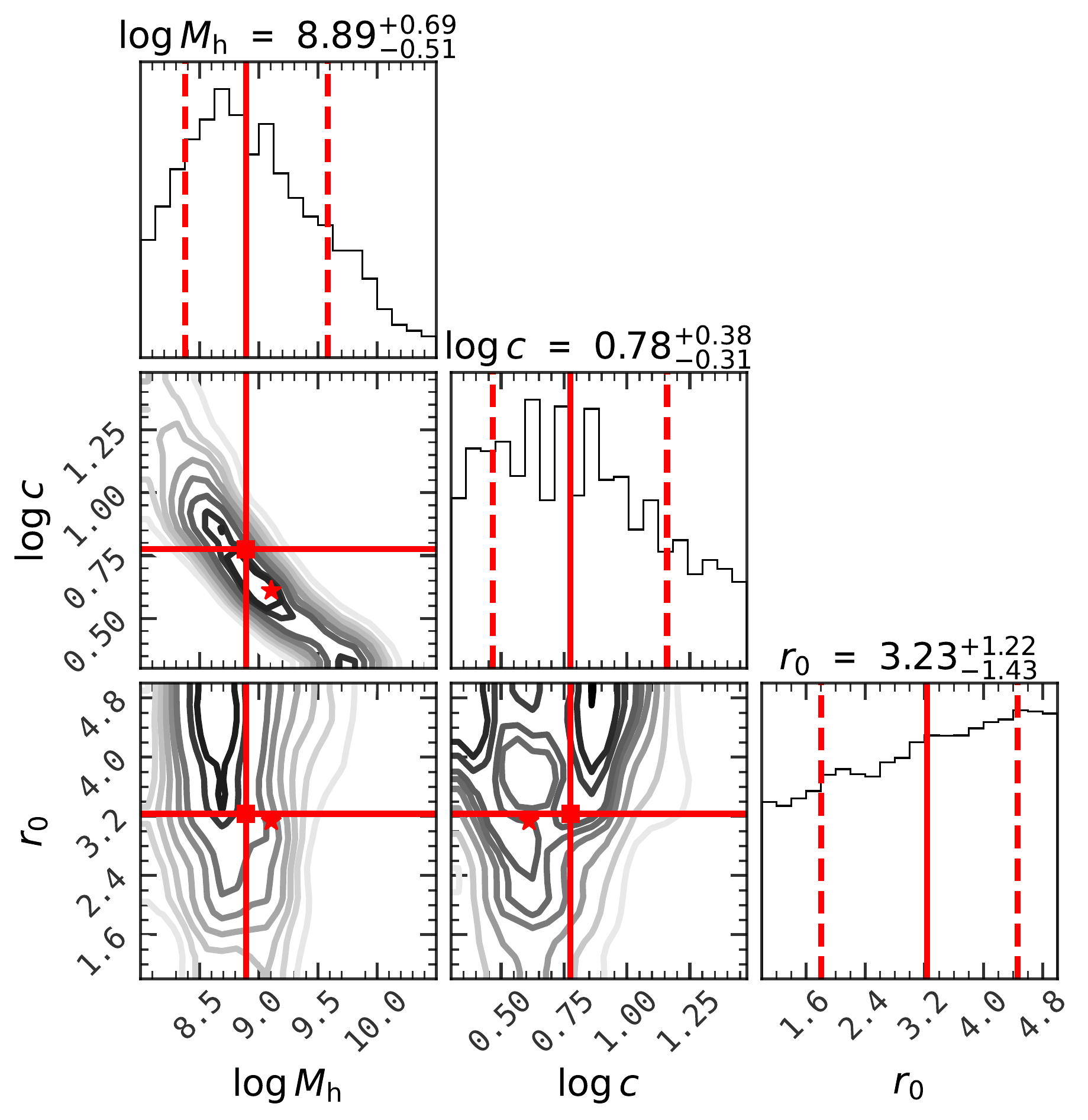}
\caption{
    Posterior distributions of the model parameters (halo mass $\Mh$, halo concentration $c$, and the scale radius $r_0$ for the initial star-cluster distribution), assuming an NFW host halo.
    The red lines indicate the median and the 16th and 84th percentiles. 
    The red stars indicate the 2D projections of the 3D mode value. 
    }
    \label{fig:cornerNFW}
\end{figure}

\begin{figure*}
\centering
\includegraphics[width=0.65\textwidth]{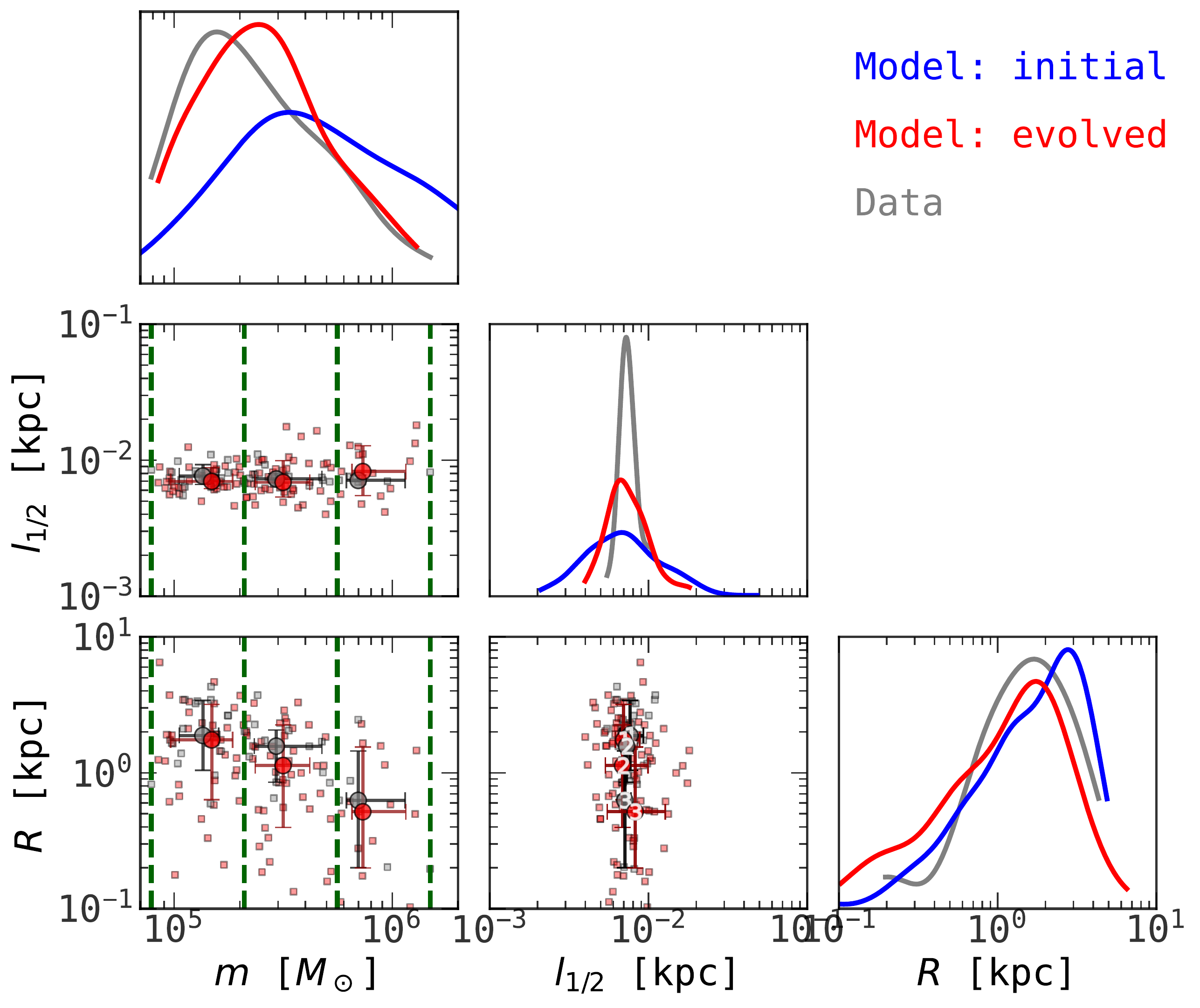}
\caption{Model realization with an NFW host halo and the mode parameters ($\Mh=10^{9.10}\Msun$, $c=4.08$, and $r_0=3.14$ kpc) compared to the data. The diagonal panels show the (individually normalized) one-point functions of star-cluster mass $m$, size $\lhalf$, and galactocentric distance $R$. The red circles stand for the median model predictions in the three mass bins (whose boundaries are indicated by the vertical dashed green lines), while the gray circles are those from observation. The error bars indicate the 16th and 84th percentiles. The numbers $i$ shown in the center of the circles in the $R-\lhalf$ plane means that this point is for the $i$th mass bin.}
\label{fig:NFWrealization}
\end{figure*}

\Fig{cornerNFW} shows the posterior distributions for NFW host halos. 
The mode values (in the 3D parameter space) are $\Mh=10^{9.1}\Msun$, $c=4$, and $r_0=3.1$ kpc, as indicated by the red stars.
The median values, together with the 16th and 84th percentiles, are $\log(\Mh/\Msun)=8.9^{+0.7}_{-0.5}$, $c = 6.0^{+8.4}_{-3.1}$, and $r_0=3.2^{+1.2}_{-1.4}$ kpc, as indicated by the red lines.
\Fig{NFWrealization} shows a model realization with the mode parameters, with 300 star clusters intially.

First, focusing on the $R$-$m$ plane of \Fig{NFWrealization}, there is a clear trend of mass segregation in the model realization, very similar to that observed. 
The parameter space that can give rise to such a prominent mass segregation is actually rather limited: a halo significantly more massive than $10^{9.5}\Msun$ can hardly reproduce the slope and the small distances of the most massive GCs, irrespective of how $c$ or $r_0$ is varied (see e.g., Appendix \ref{app:interpolation}). 

Second, the evolved GC size distribution and the size-mass relation are reproduced fairly well: 
note that the initial GC size distribution is quite broad, but the evolution shrinks the size distribution to better match that observed.  
Related, the evolved GC size-mass relation is almost flat, as observed, while the initial one has a slope of 0.24.
These two trends are largely because the low-mass GCs expand due to two-body relaxation and tidal interactions while the massive ones are almost intact, as discussed in \Fig{size-mass}.  

Third, the GC mass distribution evolves from an initial broad one towards a narrower distribution in better agreement with what is observed. 
This is partly because of the depletion of the lowest-mass clusters ($m\la 10^{4.5}\Msun$), but also partly because some of the most massive clusters ($m\ga 10^6\Msun$) have sunk to the center of the system to contribute to the formation of a nuclear star cluster and thus not taken into account here. 
We will discuss this further in \se{discussion}. 

As can be seen from the posteriors in \Fig{cornerNFW}, there is an anti-correlation between halo concentration $c$ and halo mass $\Mh$.
This degeneracy is driven by the mass segregation, which can only be achieved with an appropriate amount of DF -- 
overly strong DF would result in orbital decay that is too fast, such that all the massive GCs sink to the center, forming a stellar nucleus instead of a continuous radial mass gradient; overly weak DF would have no effect. 
Ignoring the Coulomb logarithm, the strength of DF at a radius $r$ can be estimated with the quantity $r\rho(r)/M(r)$, as can be seen from \eq{DF}, where the DF acceleration $a_{\rm DF}$ scales linearly with the local density $\rho(r)$ and inversely with the velocity squared, $V^2\sim G M(r)/r$.
For NFW profiles, it is easy to show that this quantity increases with increasing halo mass or concentration, for the radius range of interest ($r\la5\kpc$). 
This is not the case for a Burkert profile, as will be discussed shortly in \se{BurkertResult}.  

At the posterior median value, the halo mass of $\Mh\sim10^{9}\Msun$ corresponds to a $\Ms/\Mh$ ratio of $\sim 0.1$, much higher than that of normal galaxies.
Also interestingly, the concentration is much lower than the cosmological average values. 
The expected halo concentration is $\sim 25$ \citep{Dutton14} for a halo mass of $\Mh\sim 10^9\Msun$, $\sim3\sigma$ higher than the posterior median.
We discuss the implications of these findings in \se{GalHaloConnection}.

The initial star-cluster distribution, with a scale distance of $r_0\sim 3\kpc$, is more extended than the present-day smooth stellar distribution of UDG1, which has an effective radius of $2\kpc$. 
This may provide clues for understanding star cluster formation.
One scenario is that the star clusters may have formed {\it ex situ} and been brought in by satellite galaxies, which have since then been disrupted in UDG1 and released their star clusters.
The other scenario, perhaps a more natural one given the similar colors of the GCs \citep{Danieli22}, is that the clusters may have formed {\it in situ} but in an extended configuration or with high velocity dispersion, e.g., during collisions of high-redshift gas clouds that belong to different satellite galaxies \citep{Silk17,Dokkum22}. 
\\

\subsection{Burkert halo}\label{sec:BurkertResult}

\begin{figure}	
\centering
\includegraphics[width=\columnwidth]{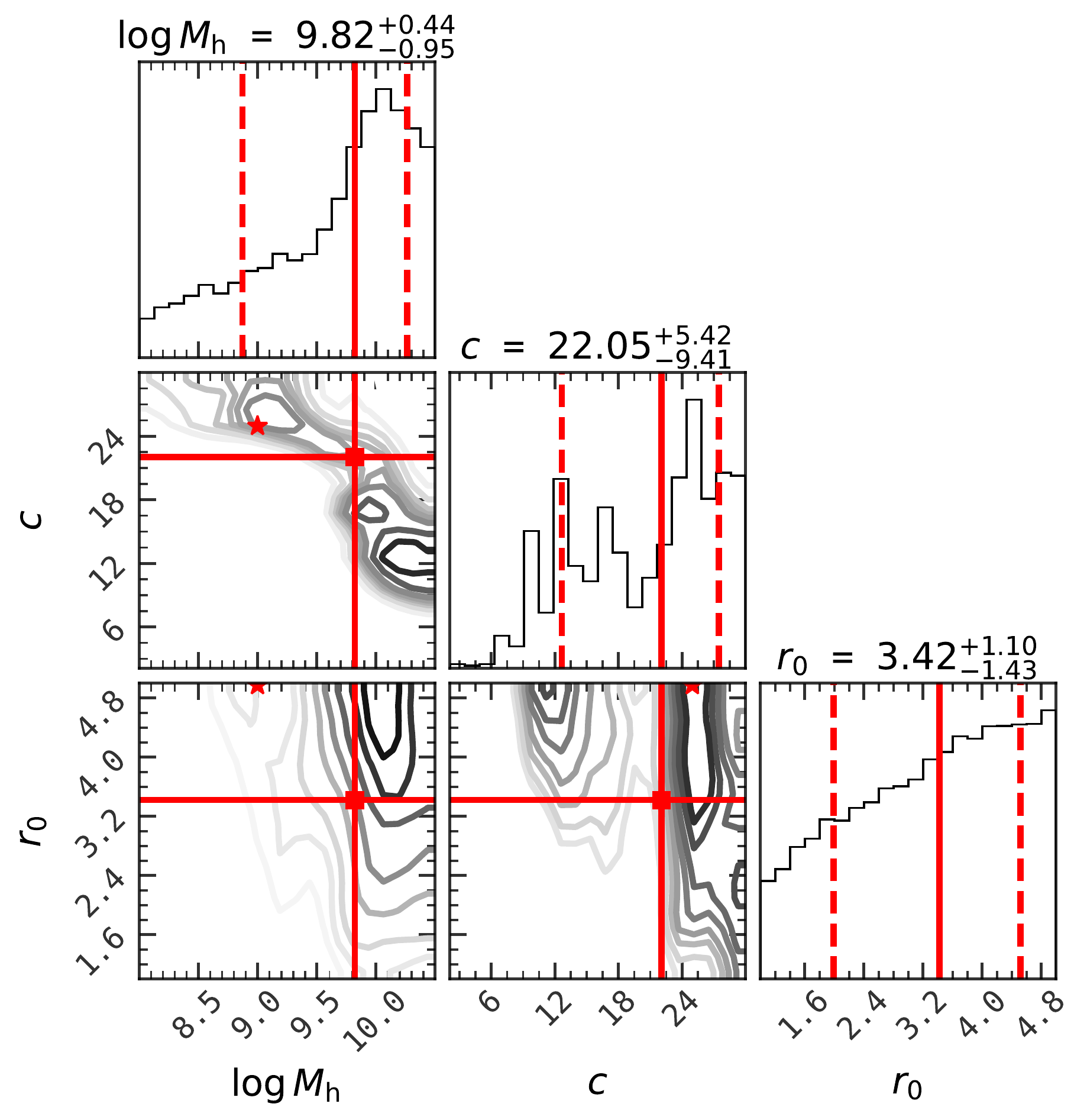}
\caption{The same as \Fig{cornerNFW}, but assuming a Burkert host halo.}
\label{fig:cornerBurkert}
\end{figure}

\Fig{cornerBurkert} shows the posterior distributions with Burkert halos. 
The results are overall similar to those with an NFW halo, but with subtle, interesting differences. 
The mode values are $\Mh=10^{9.0}\Msun$, $c=25$ ($c_{-2}=16$)\footnote{Halo concentration usually refers to $c_{-2} = \rv/r_{-2}$, with $r_{-2}$ the radius at which the logarithmic density slope is -2. The NFW scale radius $\rs$ is the same as $r_{-2}$, but the Burkert scale radius $\rs (= \rv / c)$ is $r_{-2}/1.52$, so the Burkert concentration $c$ quoted here is 1.52 times the $c_{-2}$ as commonly reported in cosmological concentration-mass-redshift relations.}, and $r_0=5.0$ kpc, and the medians with the 16th and 84th percentiles are $\log(\Mh/M_\odot)=9.8^{+0.4}_{-1.0}$, $c= 22^{+5}_{-9}$ ($c_{-2}=14^{+4}_{-6}$), and $r_0=3.4^{+1.1}_{-1.4}$ kpc.The mode halo mass is similar to that of the NFW case, both of which leaving the galaxy in the relatively DM-poor territory with respect to the stellar-mass-halo-mass relations \citep[e.g.,][]{Behroozi13}. The concentration is significantly higher, within 1 $\sigma$ of cosmological concentration-mass relations \citep[e.g.,][]{Dutton14}, but is still on the lower side for its halo mass. The median halo mass is higher, making the galaxy less extreme in terms of the stellar-to-halo-mass ratio.

\begin{figure*}
\centering
\includegraphics[width=0.65\textwidth]{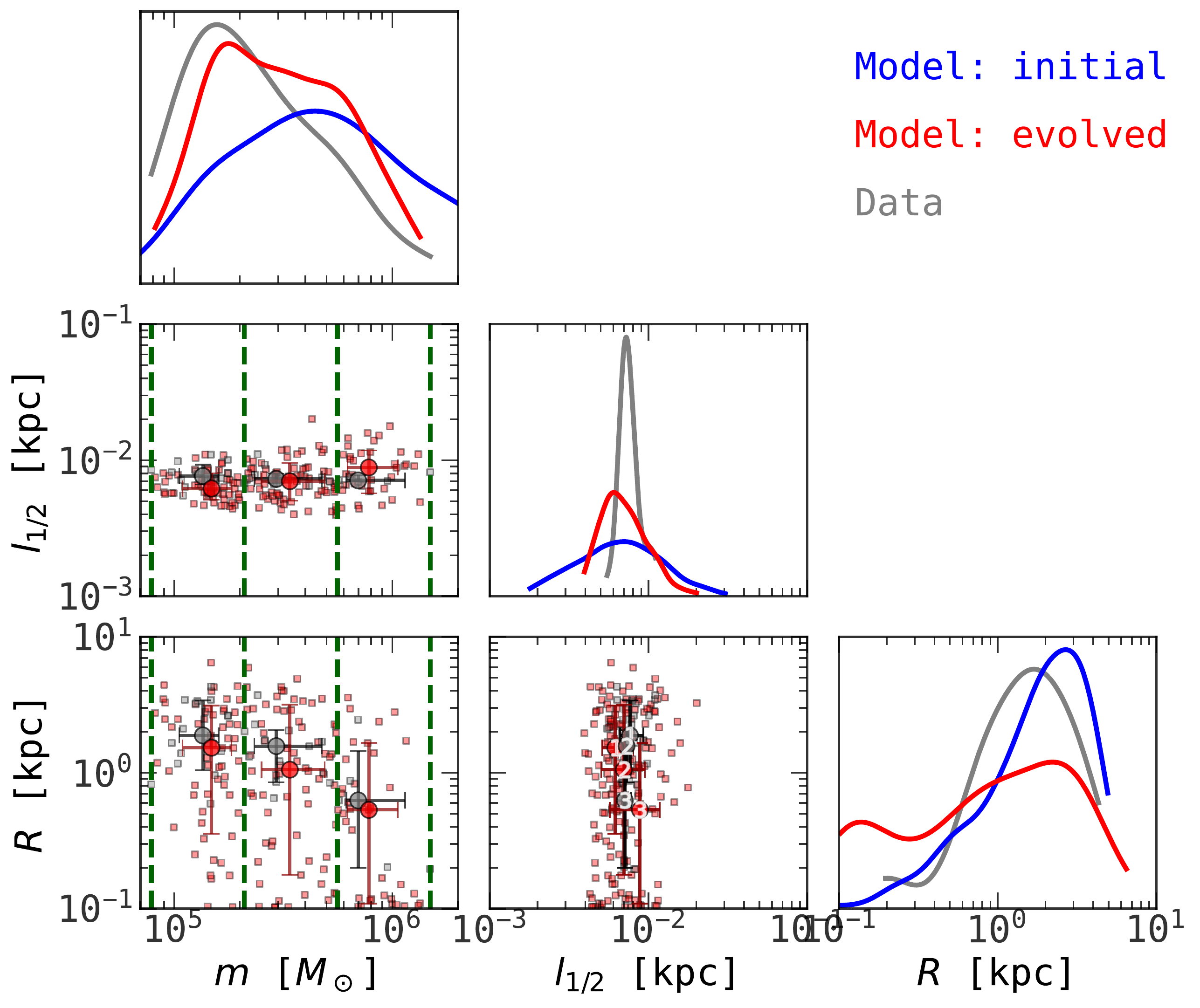}
\caption{The same as \Fig{NFWrealization}, but for the best-fit Burkert halo ($\Mh=10^{9.0}\Msun$, $c=25$ ($c_{-2}=16.4$), and $r_0=5.0$ kpc.}
\label{fig:Burkert-realization}
\end{figure*}

For the Burkert halo, the degeneracy between concentration and halo mass is weaker.  
Again, this can be understood using the proxy of DF strength, $\rho(r)/\Vc^2(r)$ --
while in the NFW case, increasing $c$ and $\Mh$ can both increase this quantity for the radius range of interest ($r\la5\kpc$), it is no longer the case with a Burkert profile. 
Instead, the $\rho(r)$ change when varying $c$ almost exactly cancels that in $\Vc^2$.
This is also why the constraining power on halo concentration is rather weak.  
For the initial scale distance $r_0$, we also obtain a median value that is larger than the effective radius of UDG1, so the same formation scenarios could be hypothesized. 

Similarly, we generate a model realization of 300 GCs with the mode parameters of the Burkert halo, and as shown in \Fig{Burkert-realization}, it also reproduces most aspects of the data. 
Hence, either a cuspy halo or a cored halo can more or less reproduce the observed GC statistics.
There is a weak but noticeable difference, that more GCs can reach smaller distances in a Burkert host: in the $R$-$m$ plane, very few model GCs with $m\sim10^5\Msun$ populate the region of $R\la1\kpc$ in the NFW host, but here there is a more significant low-$R$ tail. 
The same trend was actually also seen in \cite{Bar22}, which adopted a simpler model and ignored the details of GC evolution.  
The most obvious difference that the different profile shapes can cause is actually the fraction of GCs that reach the center of the host galaxy and form a nuclear star cluster. 
We will discuss this further in \se{discussion}. 

\subsection{Comparison with scaling relations}\label{sec:GalHaloConnection}

\begin{figure}
\centering
\includegraphics[width=\columnwidth]{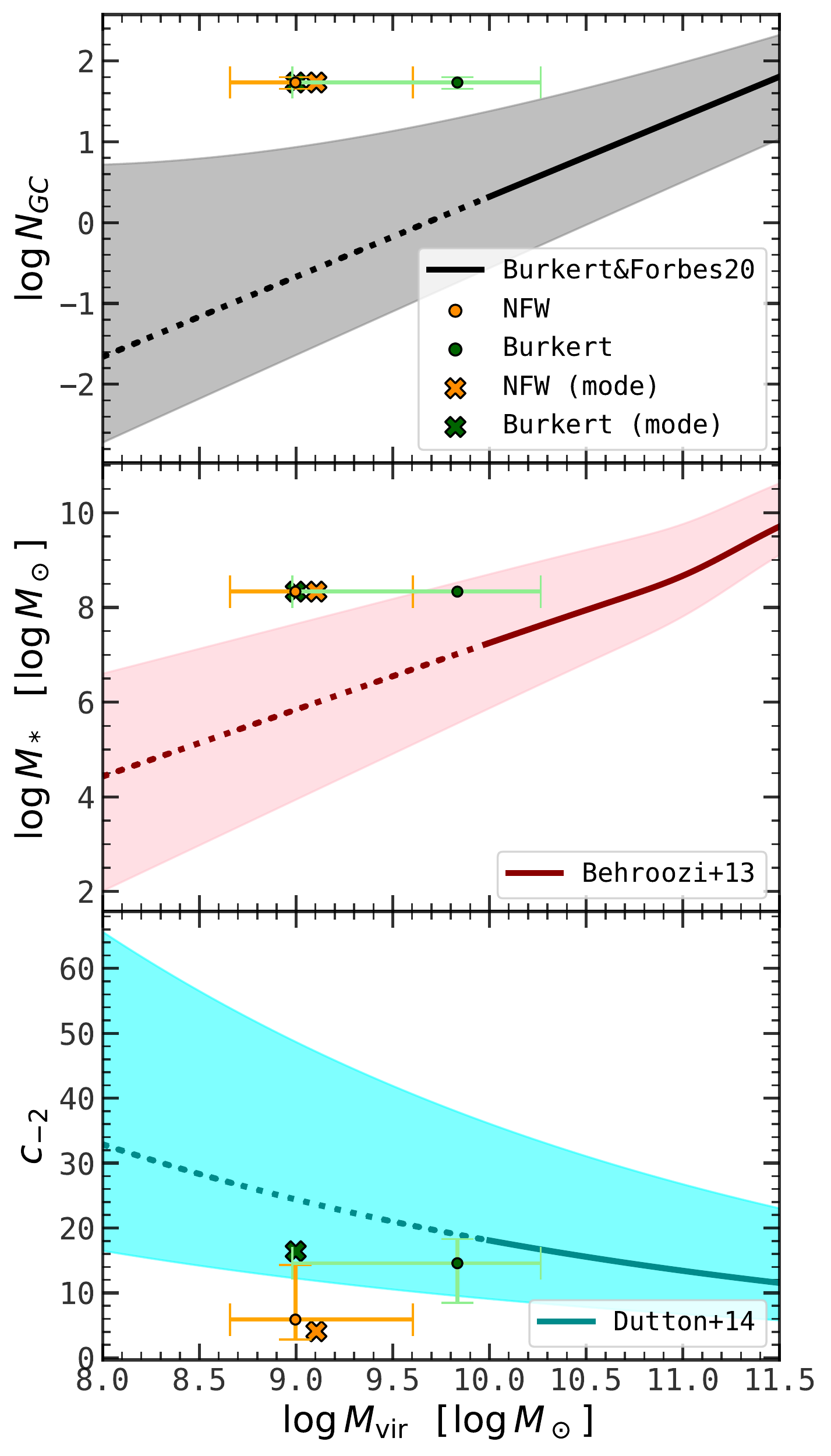}
    \caption{
    NGC5846-UDG1, with its DM halo constrained with the GC statistics, in comparison with the empirical scaling relations of GC number $N_{\rm GC}$ versus virial mass $\Mv$ ({\it upper}), stellar mass $\Ms$ versus virial mass $\Mv$ ({\it middle}), and halo concentration $c$ versus virial mass $\Mv$ ({\it bottom}). 
    The orange/green circles with error bars represent the medians with the 16th and 84th percentiles, and the crosses of corresponding colors represent the mode values, assuming NFW/Burkert halo.  
    The lines stand for the median relations from  \cite{BF20}, \cite{Behroozi13}, and \cite{Dutton14}. 
    The dashed parts of the lines indicate extrapolations to a lower mass range than that of the observational or simulation samples from which these relations are extracted. 
    The grey band in the upper panel represents the full scatter of the observational sample as in \cite{BF20}. 
    The red band in the middle panel indicates the 1$\sigma$ scatter in $\log\Ms$ at fixed virial mass, as constrained using dwarf satellites in the ELVES sample, from \citet{Danieli22b}. 
    The cyan band indicates the 1$\sigma$ scatter of $0.3$ dex, assuming log-normal distributions of $c$ at fixed virial masses. 
    UDG1 is an outlier to these scaling laws by $\sim 2-3\sigma$. 
    }
    \label{fig:relation}
\end{figure}

With the aforementioned halo-profile constraints, UDG1 is an outlier in several scaling relations of galaxy-halo connection and halo structure.  
First, for more massive galaxies, the abundance of GCs, $N_{\rm GC}$, is an excellent indicator of the host virial mass \citep{Harris17}: a simple linear relation, $\Mv = 5\times10^9\Msun \times N_{\rm GC}$, fits the observational median for almost 5 decades in halo mass from $\Mv\sim10^{10}\Msun$ to $10^{15}\Msun$.
The scatter of this relation increases towards the lower-mass end, $\sigma_{\log N_{\rm GC}}\propto \Mv^{-1/2}$, and is approximately 0.31 dex at $\Mv=10^{11}\Msun$ \citep{BF20}.
This relation has been widely used as a halo-mass estimator \citep[e.g.,][]{Forbes21} and was the basis of the hypothesis that some of the most GC-abundant UDGs are failed $L^*$ galaxies \citep[e.g.,][]{Dokkum16}. 
If we assume for simplicity that the scatter in virial mass at a fixed GC abundance is the same as that in the GC abundance at fixed mass, i.e., $\sigma_{\log\Mv}=0.31$ at $N_{\rm GC}=20$ and $\sigma_{\log \Mv}\propto N_{\rm GC}^{-1/2}$, then a galaxy with 50 GCs is expected to have a virial mass of $\Mv = 10^{10.7\pm 0.2}\Msun$.  
Hence, according to our halo-mass estimates, UDG1, with a virial mass of approximately $\Mv\sim10^{8.6-10.2}\Msun$, is a dramatic outlier to this empirical $N_{\rm GC}$-$\Mv$ relation (as extrapolated to the lower mass range) by several $\sigma$. 
We illustrate this in the upper panel of \Fig{relation}. Here, the grey band actually represents the full width of the distribution of the observational sample compiled by \citet{BF20} -- despite the increase of the scatter at the low-mass end, UDG1 is still an outlier. 

However, the $N_{\rm GC}$-$\Mv$ relation is based on massive galaxies, so the extrapolation to the low-mass end ($\Mv\ga10^{10}\Msun$) is ungrounded, and the scatter of the relation may contain systematics with morphology.
In fact, as \citeauthor{BF20} already noticed, in their effort of explaining this relation with halo merger trees, the relation must flatten at $\Mv\la10^{10}\Msun$ or $N_{\rm GC}\la100$, which is exactly the regime of GC-rich UDGs.
This flattenning is supported by the observational sample of dwarf galaxies whose virial masses are individually constrained with gas kinematics \citet{Forbes18}, as represented by the grey band in \Fig{relation}.
Our virial-mass estimate of UDG1 is qualitatively in line with the flattening of the $N_{\rm GC}$-$\Mv$ relation at the low-mass end, but more extreme, highlighting the danger of naively inverting the relation to infer virial mass with the number of GCs. 

Second, UDG1 is also an outlier to the stellar-mass-total-mass relation from abundance matching especially if assuming an NFW halo, as illustrated in the middle panel of \Fig{relation}.
For comparison, we have chosen the median relation as in \cite{Behroozi13}, and recent estimate of the low-mass-end scatter using the dwarf satellites in the ELVES sample \citet{Danieli22b}. 
We caution that despite being intensively studied, the low-mass end of the relation remains highly uncertain, and different assumptions lead to different results \citep[see e.g.,][and the references therein]{Danieli22b}.
Our particular choice here is among the most flat for the low-mass-end median slope and among the largest in the scatter -- even with these conservative choices, UDG1 is a $\sim2\sigma$ outlier if assuming an NFW halo. 

Third, as the bottom panel of \Fig{relation} shows, UDG1 stands out with respect to the concentration-mass relation of DM halos if the halo density profile is NFW.
For comparison, we have shown a median $c_{-2}$-$\Mv$ relation from cosmological $N$-body simulations \citep{Dutton14}, with a constant scatter of $0.3$ dex assuming log-normal $c_{-2}$ distribution at fixed $\Mv$ \citep{Diemer15,Benson20}.\footnote{The scatter in principle varies with mass and the selection of halos based on whether they are relaxed, and 0.3 dex is a ballpark estimate.}
Obviously, Assuming an NFW halo, the UDG1 halo is significantly less concentrated than what is expected cosmologically for its mass. 
The concentration is within 1$\sigma$ of the concentration-mass relation if the halo has a Burkert profile, but is still lower than the median relation.
Overall, this is consistent with the scenario that UDG formation results from repeated supernovae feedback, which makes both the halo less concentrated and also the stellar distribution puffy \citep[e.g.,][]{Jiang19,Freundlich20}.

In short, the UDG1 DM halo stands out as a $\sim$2$\sigma$ outlier compared to all the aforementioned scaling relations, especially when assuming an NFW profile. 
It is in line with the understanding that there is huge scatter in these relations at the low-mass end, and warns us against generalizing these relations to extreme galaxies and using them as virial mass estimators. 
We caution that our halo mass estimates for UDG1 is based on the assumption of a static host halo, whereas in reality the UDG1 halo might be a satellite of the galaxy group NGC5846, and thus have been environmentally processed. It may also have internally-driven evolution due to supernovae feedback. 
To consider the host halo of UDG1 as a subhalo evolving in mass and structure is beyond the scope of this work, but it is reasonable to speculate that the peak virial mass of the system in the past is higher than our estimates here, and thus brings the system closer to the empirical scaling laws. 


\section{Discussion}\label{sec:discussion}

In this section, we first discuss a few observational implications, including the line-of-sight (LOS) velocity dispersion of the GCs in UDG1, the orbital eccentricity distribution of the GCs, the stripped mass fraction of the GCs and the fraction of nuclear star clusters. 
Second, we compare our model with that of the previous work of \cite{Bar22} and \cite{Modak22}. 
Finally, we comment on the simplifications in this work, and point out potential future improvements and applications of our method. 
We explore the possibility of an initial mass segregation in \app{init_segre}. 

\subsection{Velocity dispersion and orbital eccentricity of the GCs}\label{sec:kinematics}

\begin{figure*}
\centering
\includegraphics[width=\textwidth]{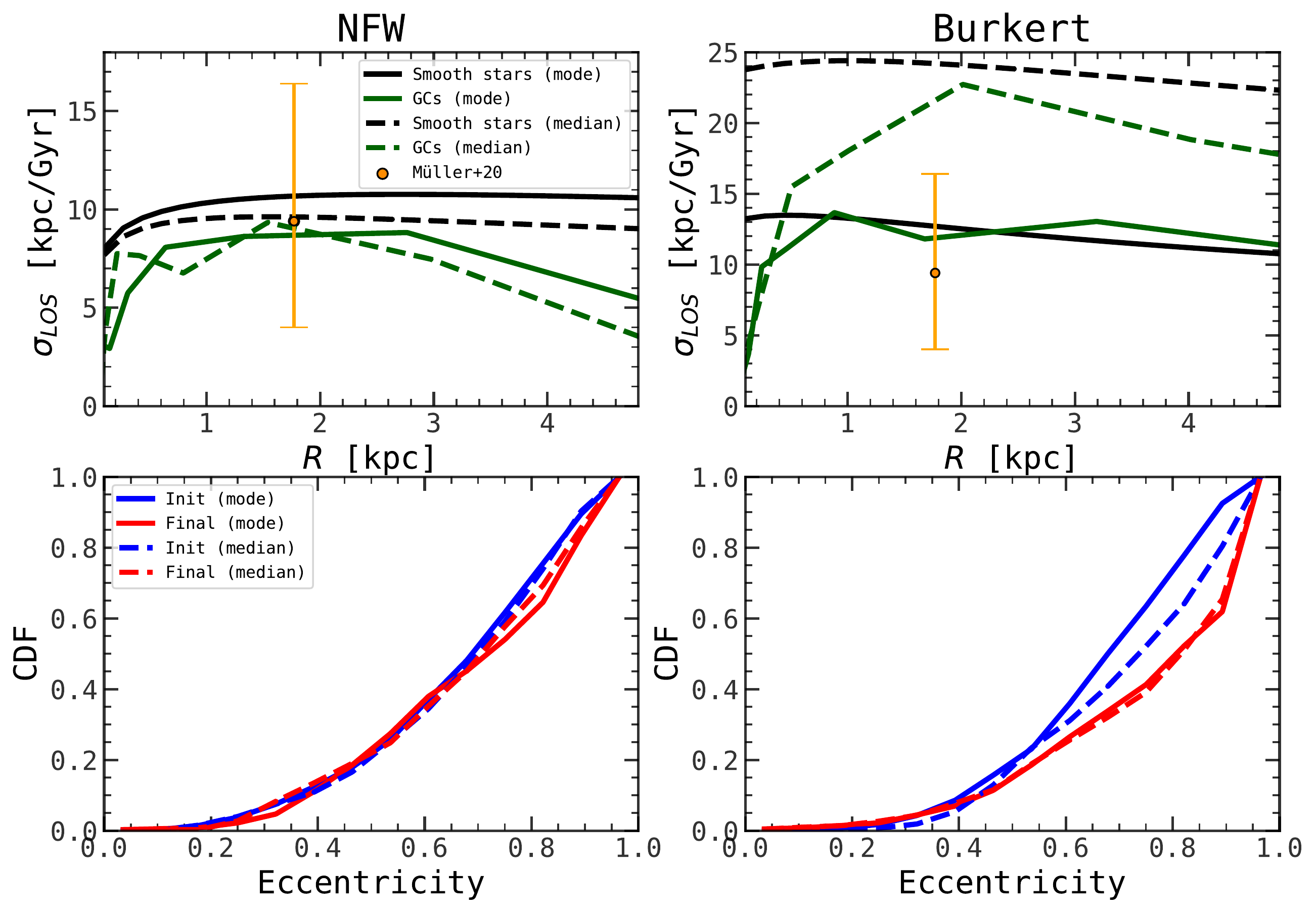}
    \caption{Upper panels: LOS velocity-dispersion profiles of the GC population in comparison with that of the smooth stars. 
    Lower panels: cumulative distributions of orbital eccentricity of the star clusters.
    The left column represent the result for NFW halo while the right column represent the result for Burkert halo. 
    The solid lines represent our best-fit model results while the dashed line represent our median model results. In the upper panels, black lines stand for smooth stars and green lines stand for evolved GCs. In the lower panels, red lines stand for the final stage and blue lines stand for the initial distribution, and the line style differentiate the mode and median results, respectively.
    The orange circle with error bar is inferred from the observed kinematics, based on about one fifth of all the $\sim 50$ member GCs \citep{Muller20}. }
    \label{fig:vdisp}
\end{figure*}

In the upper panels of \Fig{vdisp}, we present the LOS velocity-dispersion profile of the evolved GCs in halos of the best-fit parameters. 
The dispersion profiles in the Burkert halo and in the NFW halo are similar, and, in comparison with that of smooth stars, which reflect the equilibrium kinematics of the host potential, both show a significant decrease, more so at smaller $R$.  
This is another manifestation of DF, besides mass segregation.
Observationally, \citet{Muller20} have measured the velocities of 11 of the member GCs of UDG1, and inferred a dispersion of $\sigma = 9.4^{+7.0}_{-5.4}\kms$ at $R\approx1.8\kpc$ (i.e., the average distance of the 11 GCs to the galaxy center), assuming a simple pressure-supported spherical system.
Our model predictions agree with this measurement.
In fact, a closer examination of the posterior distribution of the GC velocity dispersion of \citet{Muller20} reveals that the mode value is approximately $8\kms$, almost exactly on top of our NFW model results. 

Interestingly however, \citet{Forbes21} measured the velocities of the smooth stellar distribution of UDG1 using KCWI on the Keck telescope, and found a rather high value of the smooth-star velocity dispersion of $\sigma_\star=17\pm2\kms$, higher than the equilibrium value ($\sigma\sim11\kms$) of a low-mass ($\Mv\sim10^9\Msun$) and extremely low-concentration system as advocated by our best-fit NFW cases, and more consistent with a virial mass of $\Mv\sim10^{9.5-10}\Msun$ and of normal concentration as in the Burkert cases.
In none of the models explored here, we reproduce such a large difference between the GC dispersion and the smooth-star kinematics.
We opt not to dive into the factors that may reconcile the tension, such as oversimplifications in our model or non-equilibrium of the stellar distribution. 
Instead, we can see that these two observational studies together \citep{Muller20,Forbes21} present a qualitatively similar picture as what our model reveals here, i.e., the GCs of UDG1 have smaller velocity dispersion than the smooth stars, indicative of DF, and the virial mass of UDG1 is lower than what is expected from the scaling laws.  

The different GC dynamics with the two halo profiles may result in different orbital eccentricities of the GCs. As shown in the lower panels of \Fig{vdisp}, while the orbital eccentricity distributions are initially similar (which is by construction, because we assumed isotropic velocity distribution in both cases), evolution in the Burkert host makes the orbits slightly more eccentric. 
This effect actually only operates on the GCs whose initial apocenter is larger than the core radius and whose initial pericenter is below the core radius. 
Such GCs experience dynamical friction only at the apocenter but not at the pericenter because of core-stalling -- this makes the orbits more radial. 
Note that dynamical friction will otherwise not affect orbital eccentricity, a counter-intuitive behavior already discussed in \citet{vdB99}, because eccentricity decreases near peri-center but increases again near apocenter.

\subsection{Nucleated ultra-diffuse galaxies}\label{sec:nucleated}

A significant fraction of UDGs are nucleated, in the sense that they feature a compact stellar distribution at or near the geometric center of the system \citep{Lim18,Greco18,Iodice20,Marleau21}.
The compact stellar source, also known as the nuclear star cluster (NSC), is more compact than a stellar bulge as in an early-type galaxy and is more massive than a typical GC -- imaging samples can be found at \citet[][Fig.3]{Lim18}.
The fraction of UDGs that are nucleated is approximately 30-40\% in nearby galaxy clusters, and seem to show an environment dependence such that the fraction is higher in the densest region and decreases towards the outskirts of the galaxy cluster \citep{Lim18}. 

It it natural to attribute the formation of an NSC to the coalescence of the GCs which have lost their orbital angular momentum completely due to DF and sunk to the center.
If this is the case, we can expect that different DM profiles, as well as different initial GC distributions, can determine the nucleatedness of a UDG and the mass of the NSC.
\citet{Modak22} already showed that NSCs only form in cuspy halos and almost never form in a cored halo.
Here we revisit this scenario using our model, which is more refined in terms of GC evolution compared to the previous work. 
We emphasize that this experiment is for GC-rich dwarfs in general, no longer aimed at reproducing UDG1. 
We quantify the nucleatedness of the resulting system using the mass fraction of the nucleus, $f_{\rm nucleus}$, defined as the total mass of the GCs that settle to $r<0.1\kpc$, divided by the total mass of all the GCs. 

\begin{figure*}
\centering
\includegraphics[width=1.0\textwidth]{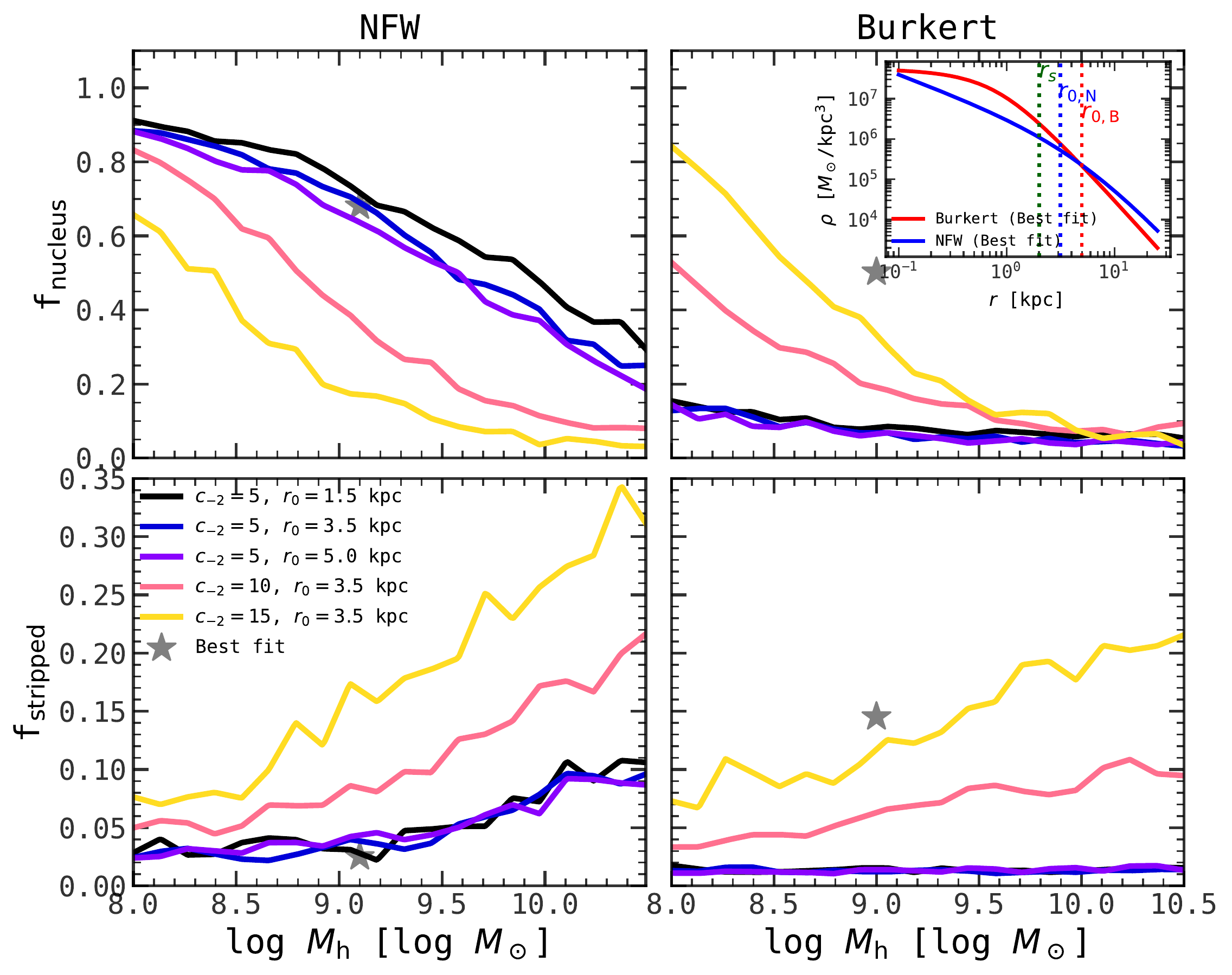}
    \caption{
    Upper panels: The mass fraction of nuclear star cluster (the total mass of the GCs that have sunk to the center of the galaxy due to dynamical friction, divided by the total mass of all the GCs in the galaxy) as a function of halo mass. Lower panels: The stripped fraction of GCs (the total mass difference between GCs in beginning and in final stage, divided by the total mass of all the GCs in the galaxy) as a function of halo mass. The left and right column show the results assuming cuspy (NFW) and cored (Burkert) profiles, respectively. Different colors represent different combinations of the halo concentrations ($c$) and the scale length ($r_0$) of the initial GC distribution, as indicated. The inset panel show the density profile of NFW halo (red solid line) and Burkert halo (blue solid line) with best fit parameters. The vertical green dotted line represent effective radius of the galaxy while the blue dotted line and red dotted line stand for the best fit initial GC scale radius for NFW case and Burkert case, respectively.}
    \label{fig:nuGC_frac}
\end{figure*}

\Fig{nuGC_frac} shows $f_{\rm nucleus}$ as a function of halo mass $\Mh$, halo concentration $c$, and the initial Hernquist scale of the GC distribution $r_0$.
Here, a cuspy NFW profile and a cored Burkert profile exhibit a dramatic difference, in the sense that the nucleus fraction in a cored halo is usually much lower (except for the combination of the lowest $\Mh$, highest $c$, and largest $r_0$, which happens to be our best-fit case), whereas that of a cuspy halo is usually much higher, especially when the halo mass is below $\Mh \sim 10^{9.5}\Msun$. 
The generally low nucleus fractions of a cored halo is due to the stalling effect, which prevents the GCs from dropping deeper \citep[see][for a thorough discussion]{Banik22} and is empirically captured by the \citet{Petts15} prescription adopted here.  
However, in our best-fit Burkert profile which has a fairly high concentration, while there is a flat core in the very center, the density is still steeply rising towards the center in the radius range where most GCs are. 
The inset of \Fig{nuGC_frac} compares the best-fit NFW and Burkert profiles:
we can see clearly that the Burkert profile is even steeper than the NFW counterpart at the initial Hernquist scale radius for the GC distribution.
It is this steep density slope at the regime where the GCs populate that gives rise to sufficiently strong DF and therefore mass segregation. 
Otherwise, if the core radius is larger than where the star clusters are, there will be no mass-segregation trend, and, for the same reason, no NSC forming.

There are trends of the nucleus fraction with the model parameters. 
First, larger halo mass leads to smaller $f_{\rm nucleus}$. 
This is because of the dependence of the DF strength on the mass ratio between the GC and the host. 
Second and intuitively, larger scale length leads to smaller $f_{\rm nucleus}$ , since if GCs start out at large distances, they need stronger DF or longer time to sink to the center. 
Third, as to the concentration dependence, for NFW halos, a higher concentration leads to a lower $f_{\rm nucleus}$, at fixed halo mass. 
This, again, can be attributed to the competing effects on DF from $c$ and $\Mh$, as we discussed in the context of the $c_{-2}$-$\Mh$ degeneracy of a cuspy halo in \se{NFWresult}.
For Burkert halos, the $c_{-2}$-trend is vague, and can similarly be attributed to the null effect of varying $c_{-2}$ on the DF acceleration as we argued in \se{BurkertResult}, unless the halo mass is sufficiently low such that a higher $c_{-2}$ makes the regime of steep density slope overlapping with the GCs. 
The difference in the nucleus fraction between the cuspy and cored cases implies the possibility of using NSCs to infer the DM distribution of UDGs, and could be further explored in future statistical studies.

\subsection{Stripped fraction of GCs}\label{sec:stripped}

A higher host-halo density will lead to more tidal stripping of star clusters, as can be expected from \eq{lt}. Hence, naturally, a positive correlation exists between halo concentration and the stripped mass fraction of the GCs, as shown in the lower panels of \Fig{nuGC_frac}. 
For the same reason, when concentration is fixed, a more cuspy halo results in more stripped mass from the star clusters -- this can be seen by comparing the NFW and Burkert results in the lower panels of \Fig{nuGC_frac}.

However, if we limit the scope of comparison to the halos that can produce the observed mass segregation, we get the seemingly counter-intuitive result that, in the best-fit NFW case, the stripped fraction of the GCs is marginal, whereas in the best-fit Burkert model, the GCs have contributed a fairly large fraction ($\ga 10\%$) of their masses to the smooth stars.
This is simply because the best-fit Burkert halo is actually denser in the regime where most of the GCs exist ($r\la r_0$), as shown by the inset of \Fig{nuGC_frac}.

Note that the star clusters which have been completely disrupted also deposit their mass to the smooth-star reservoir. 
It is an interesting question whether ultra-diffuse galaxies obtain a significant (or even dominant) fraction of their smooth stellar mass from stripping as well as disrupted star clusters \citep[e.g.,][]{Danieli22b}. 
However, since the fraction of completely dissolved star clusters depends sensitively on the $\Mmin$ parameter of the ICMF, and $\Mmin$ is not a parameter that we have sufficient constraining power based on radial mass segregation, we cannot make conclusive argument on this scenario.

\subsection{Comparison with previous studies}\label{sec:comparison}

\begin{figure*}
\centering
\includegraphics[width=0.8\textwidth]{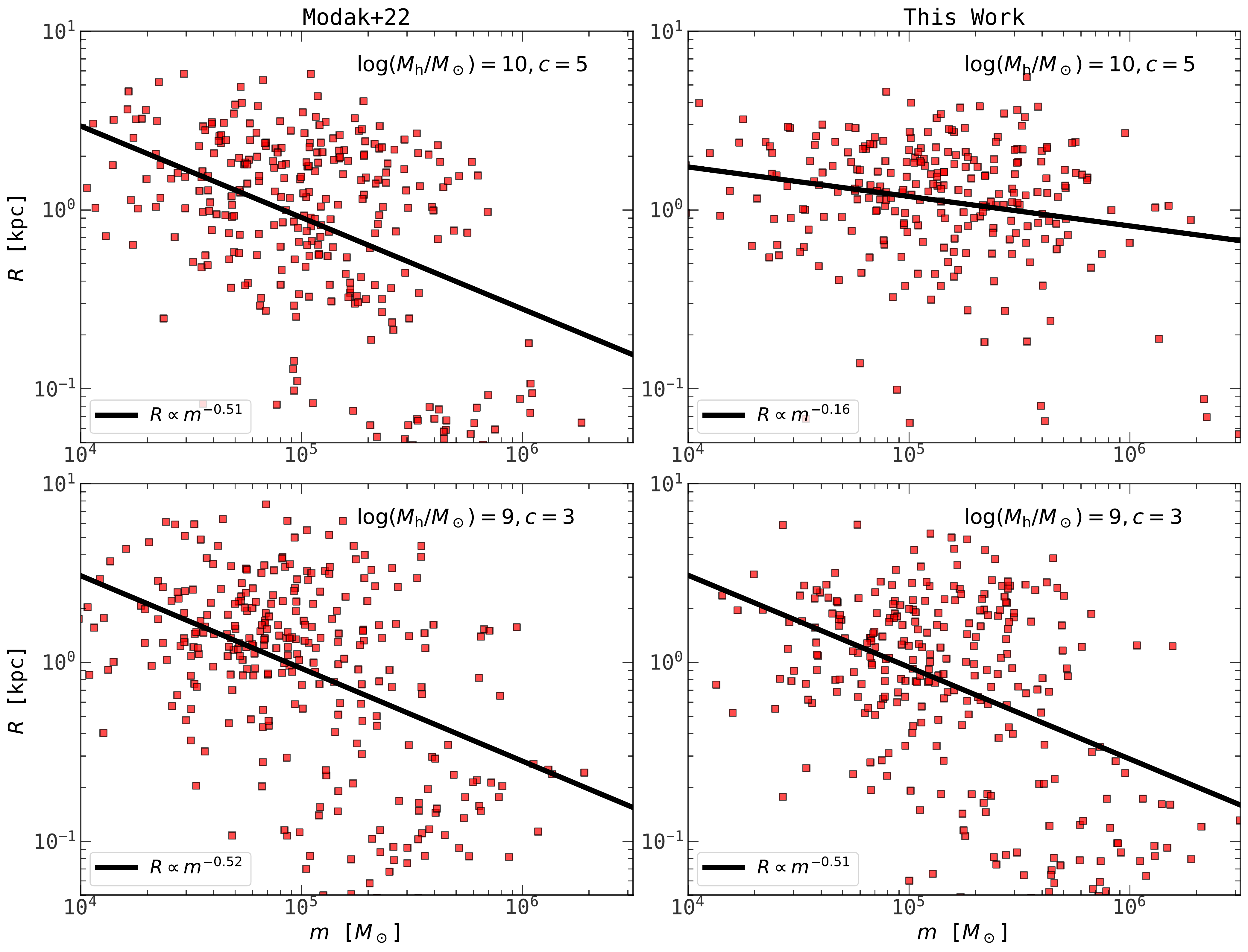}
    \caption{Illustration of the impact of the star-cluster evolution model on the halo mass at which GC mass segregation occurs: projected galacto-centric distance versus GC mass, in host halos of given mass and concentration, as quoted, obtained using the \citet{Modak22} model ({\it left}) and the model presented in this work ({\it right}) with the same initialization scheme. To make the results statistically robust, 20 random realizations of 50 initial star clusters are included.  
    Linear regression results are overplotted to gauge the strength of the mass segregation. 
    {\it Upper panels:} with a relateively massive halo ($\Mh=10^{10}\Msun$), our model produces marginal mass segregation, while the \citeauthor{Modak22} model, featuring orbit-independent mass loss and neglecting star-cluster size in the dynamical-friction treatment, can already produce fairly strong mass segregation. 
    {\it Lower panels:} with a lower halo mass of $\Mh\sim 10^9\Msun$ and a low concentration, both models can produce rather strong mass segregation.
    }
    \label{fig:realization_comparison}
\end{figure*}

\cite{Bar22} also studied the mass segregation of GCs in UDG1 using a semi-analytical model of similar nature to ours. 
\cite{Modak22} studied GC statistics in a more compact dwarf galaxy but adopted a very similar model to that of \citeauthor{Bar22}.
There are a couple of major improvements here in our approach. 

First, the previous studies did not trace the mass and structural evolution of GCs along the orbits, but instead adopted, e.g., as in \citeauthor{Bar22}, a simple empirical model of GC mass loss, $m(t)=m(0)\left(1-\delta t/t_0\right)$ with $\delta=0.3$, $t_0=10$ Gyr, for all the GCs. 
That is, the mass evolution is linear with time, irrespective of the local tidal field, and there is no structural evolution.
\citeauthor{Modak22} makes the mass evolution exponential, but the other aspects remain the same. 
Related, the GC mass distribution in these works were manually set to match that observed.
As discussed in \se{model}, the mass and structural evolution affects the DF strength, via the Coulomb logarithm, and thus the orbit evolution could be different if the mass or size is not properly accounted for. 
In our model, the GCs evolve self-consistently in mass and structure under the influence of tidal effects and two-body relaxation. 
Besides, the evolution starts from theoretically-motivated or observationally-motivated initial mass functions and initial structural distributions.
As such, besides the GC mass segregation signal, the evolved GC mass function and size-mass relation are also emergent predictions in our model, and, as demonstrated, they both agree decently with the data. 

Second, the previous studies did not aim at statistically constraining the DM halo mass or structure. 
Instead, they adopted a couple of somewhat arbitrary fixed halo masses and profiles, and tested whether mass segregation could arise under these somewhat arbitrary choices. 
For example, \citeauthor{Bar22} used an NFW host with $\Mh=10^{9.78}\Msun$ and $c_{-2}=6$ and a Burkert host with $\Mh=10^{9.53}\Msun$ and $c_{-2}=15.4$.
Basically, their halo masses are approximately 0.5-1 dex higher than what we found here, and their halo concentrations happen to be in the same ballpark with our posterior mode values. 
Interestingly however, they could achieve mass segregation with these more massive halos, whereas our model cannot. 

The main factor that causes the difference lies in the star-cluster evolution -- with our model,  mass loss depends on the instantaneous internal and external conditions, such that the clusters getting closer to the host center experience stronger mass loss. This counter-balances the mass-segregation effect of DF. 
Hence, to obtain mass segregation with our model, stronger DF is needed, which translates to a lower halo mass because of the dependence of DF on the perturber-to-host mass ratio.
This is illustrated in \Fig{realization_comparison}, where we keep the host halo as well as the initialization of coordinates and velocities the same, and evolve the star clusters using the \citeauthor{Modak22} model and our model, respectively, for results shown in the left and right panels.
The mass initialization cannot be exactly the same because of the difference in the evolution prescriptions, but we have adjusted the ICMF to make the evolved mass distributions comparable. 
As can be seen, with the \citeauthor{Modak22} model, significant mass segregation can already be produced for a relatively massive NFW host halo of $\Mh = 10^{10}\Msun$, while our model produces a marginal trend for this mass.
For the lower halo mass of $\Mh = 10^{9}\Msun$, both models can produce similarly strong mass segregation.  

Another factor here is that the previous studies did not follow star-cluster structural evolution. 
As we already discussed, neglecting size results in a larger Coulomb logarithm and thus stronger DF.
In short, it is easier to get mass segregation with the \citeauthor{Modak22} and \citeauthor{Bar22} models, because the mass loss is orbit-independent and because DF is stronger when neglecting the cluster size. 
It is therefore important to model GC evolution accurately for the purpose of constraining DM distribution. 

\subsection{Simplifications and future work}

While in this work we have improved upon previous studies by introducing a self-consistent physical model of GC evolution and employing MCMC to constraint the DM halo properties, we caution that there are still oversimplifications that leave room for future improvements. 
Addressing them quantitatively is beyond the scope of current work, but here we point out qualitatively how they might affect the results and sketch ideas for future studies. 
The discussion applies not just to the specific galaxy UDG1, but to GC-rich dwarf galaxies in general.

First, we have assumed that the host potential is static over the entire evolution of the GC population. 
However, UDG1 is a satellite galaxy of the galaxy group NGC5846, and many GC-rich low-surface-brightness galaxies are members of galaxy groups or clusters. 
That is, the host halo of UDG1 may be a subhalo that has experienced significant mass loss if it had orbital pericenters sufficiently close to the center of the host group. 
In the case, UDG1 may lie closer to the scaling relations (\se{GalHaloConnection}) if its peak virial mass and the concentration at the peak mass are used in place of the present-day values. 
In fact, it has been argued that the differential tidal mass loss between the subhalo and the stellar component can produce DM-deficient dwarf galaxies \citep{Moreno22}.
Even for an isolated dwarf galaxy, the host halo is not static, but increases in mass gradually. 
It is in principle possible to parameterize the mass assembly history of the host halo or subhalo using empirical models extracted from cosmological simulations \citep[e.g.,][]{Wechsler02}.
However, this would introduce additional model parameters that need to be marginalized over, not to mention that there is significant halo-to-halo variance in the mass histories \citep{JB17} so that a certain choice of the parameterization may not  be representative. 
A more viable way of exploring statistically the effect of a dynamic host is to post-process cosmological numerical or semi-ananlytical simulations and populate simulated halos with GCs and study the GC statistics. 
We leave this idea to a future study. 

Second, there are a few other mechanisms for GC mass and structural evolution besides tidal interactions and two-body relaxation, including, among others, stellar evolution and gravothermal core-collapse.  
\cite{Lamers10} provide an empirical formula for mass loss of GCs due to stellar evolution obtained from collisional $N$-body simulations.
Following their model, GCs lose $\sim$ 25\% of their initial mass over 10 Gyr, insensitive to their initial masses. 
Therefore, this effect can simply be offset by shifting the ICMF.
The more complicated effect is the gravothermal core-collapse of GCs, which steps in after when an isothermal core is established due to two-body relaxation. 
The core contracts since it is dynamically hotter than the outer part and transports energy to the outskirts.
This makes the GC profile deviate from the EFF profile assumed in this work (or more generally, the King models), and become cuspy and resistent to tidal mass loss. 
The core-collapse timescale has been estimated to be 12-19 times the relaxation time $\tau_{\rm r}$ given in \eq{RelaxationTime} \citep{Quinlan96}, so it can be shorter than the Hubble time for the lower mass GCs ($m\la 10^{4.5}\Msun$).  

Third, if future kinematics observations can narrow down the DM profile, we can then adapt our model to constrain the other model ingredients.
For instance, while the ICMF is believed to follow the functional form of \eq{ICMF}, the power-law slope as well as the mass scales are not fully constrained and likely exhibit variation from one population to another \citep[see e.g.,][for a discussion on the power-law slope]{AG13}. 
It would be interesting to treat the ICMF parameters as free parameters, and combine the GC evolution model and the MCMC method as in this study to constrain the ICMF, to investigate, for example, whether UDGs have a unique ICMF. 


\section{Conclusion}\label{sec:conclusion}

In this work, we are motivated by a remarkable ultra-diffuse galaxy, NGC5846-UDG1, whose globular-cluster population exhibits interesting radial mass segregation, and aim to explore the possibility of reproducing the mass segregation of the GCs with dynamical friction and constraining the dark-matter content of the UDG using photometric data alone.
To this end, we have introduced a simple semi-analytical model that describes the evolution of GC populations in their host DM halo and galaxy, accounting for the effects of DF, tidal evolution, and two-body relaxation. 
We also consider educated assumptions of the initial properties of the GC progenitors, including the mass function, structure, and spatial distributions.
We forward model the GCs in a UDG1-like host potential (consisting of a DM halo and a smooth stellar distribution) to match the observed GC statisitcs from \cite{Danieli22}, and use MCMC to constrain the DM distribution (halo mass $\Mh$ and concentration $c$) of UDG1, as well as the scale radius of the initial star-cluster spatial distribution ($r_0$). 
While we have focused on UDG1, the methodology developed here is generally applicable to dwarf galaxies in equilibrium with a rich GC population. 
We summarize our methodology and the main findings as follows. 

We have shown that the orbital evolution under the influence of DF depends on the mass and structural evolution.
Our model can self-consistently evolve the mass and structure of individual GCs along their orbits, capturing the effect of a varying tidal field along an eccentric orbit around the central region of a galaxy. 
In the limit of a weak tidal field, the mass and structural evolution in our model reduces to that of the classical work of \cite{Henon65}; while in the limit when the timescale for tidal stripping is shorter than two-body relaxation, the structural evolution follows that of the empirical tidal evolution track for collisionless systems of \cite{Penarrubia10} with cored density profiles, which applies to the assumed EFF density profiles of the GCs. 
Reassuringly, over the timescale of $\sim10$ Gyr, a population of star clusters drawn from reasonable initial cluster mass functions \citep{Trujillo-Gomez19} and initial structure-mass distributions \citep{BG21} evolves in a converging manner regarding its evolved mass and size distributions -- notably, the lower-mass clusters ($m\la10^{4.5}\Msun$) expand and get dissolved more easily, whereas the most massive clusters ($m\ga 10^6\Msun$) remain largely intact, making the evolved GC mass function peak at $m\sim10^5\Msun$ and the evolved GC size-mass relation flat, as observed. 

No matter whether the density profile of UDG1 is cuspy or cored, we find that the DM halos that can give rise to the observed mass segregation are of low mass and low concentration.
In particular, with an NFW (Burkert) halo, we obtained a posterior-mode halo mass of $\Mh=10^{9.1}\Msun$ ($10^{9.0}\Msun$) and concentration of $c_{-2}\approx 4$ ($16$). 
There is a concentration-mass degeneracy (anti-correlation) in the case of an NFW profile, driven by the similar effects of increasing concentration and increasing mass on the DF strength in the central few kpc of the host potential.    
Given the stellar mass of UDG1 of $\Ms\sim10^8\Msun$, these halo-mass estimates put UDG1 in the DM-poor territory. 

In fact, UDG1 is an outlier compared to both the stellar-to-total-mass relation \citep[e.g.,][]{Behroozi13, Danieli22b} and the GC-abundance-total-mass relation \citep[e.g.,][]{Harris17,BF20}.
The latter relation is known to flatten and increase in scatter at the low-mass end, and UDG1, with our halo-mass estimates, is in qualitative agreement with this trend, though more extreme. 
This warns against using this relation naively for halo-mass estimates for UDGs. 

The estimated halo concentrations are lower than the cosmological average value expected for halos of the posterior masses. 
This lends support to the theoretical picture that UDGs populate low-concentration halos, which are puffed up by repeated supernovae outflows or environmental effects \citep[e.g.,][]{DiCintio17,Chan19,Jiang19}.  
The posterior scale distance of the initial star-cluster distribution (which is assumed to follow an Hernquist profile) is $r_0 \sim 3\kpc$. 
Hence, the star clusters were likely in a more extended configuration initially than the (present-day) smooth stars which has an effective radius of $2$ kpc. 
This may imply that the star clusters are either of {\it ex-situ} origin or formed {\it in situ} but in an extended configuration achievable via collisions of high-redshift gas clouds.

The radial mass segregation of GCs can be reproduced with either assumption of the halo profile, if we just focus on the distance-mass slope of the GCs that have not sunk to the center of the galaxy (\se{results}). 
However, if we include all the GCs including the ones that have completely lost orbital angular momentum due to DF, and consider the nuclear star cluster that form out of GC mergers at the center of the galaxy, then the cuspy NFW halo can yield massive nuclear star clusters provided that the halo mass is below $10^{10}\Msun$, whereas cored halos do not result in any significant nuclear star cluster. 
Therefore, a viable formation mechanism for nucleated UDGs \citep[e.g.][]{Lim18} is the orbital decay of GCs in a low-mass cuspy halo \citep[see also][]{Modak22}. 
As UDG1 seems to be a non-nucleated UDG, our results suggest that it is more likely to be hosted by an cored, low-mass DM halo.
This is, again, in line with the theoretical picture that UDG formation goes hand-in-hand with the core formation of DM halos due to non-adiabatic perturbations of the gravitational potential. 

Last but not the least, compared to the observationally-costly kinematics measurements, our model can reproduce the observed LOS velocity dispersion of the GCs \citep{Muller20}, and can reveal the difference between the velocity dispersion of the GCs and the smooth stellar background. 
This also manifests DF and is in qualitative agreement with what is observed \citep{Forbes21}. 

In summary, we have demonstrated with the case study of UDG1 that, as long as dwarf galaxies host a statistically significant number of GCs and the GCs form a radial mass trend, one can use a computationally efficient semi-analytical model such as the one laid out in this work to constrain the hosting DM distributions. 
This is in principle feasible with imaging data alone.
However, getting clean samples of GCs with little contamination from background galaxies or foreground stars requires deep, high-resolution imaging -- this will soon be enabled by upcoming instruments, such as the Vera C. Rubin Observatory (LSST), the Chinese Survey Space Telescope (CSST), and Nancy Grace Roman Space Telescope (WFIRST). 


\begin{acknowledgments}
FJ thank Avishai Dekel, Aaron Romanowsky, Frank van den Bosch, Hui Li, Kyle Kremer, Xiaolong Du, Ethan Nadler, and Jacob Shen for helpful general discussions. 
JL acknowledges the Tsinghua Astrophysics High-Performance Computing platform at Tsinghua University for providing computational resources for this work. 
\end{acknowledgments}

\vspace{5mm}
\software{{\tt EMCEE} \citep{Foreman-Mackey13}, \SatGen \citep{Jiang21}}


\bibliography{GCinUDG}{}
\bibliographystyle{aasjournal}


\appendix

\section{Analytics of stellar profiles and DM halo profiles}\label{sec:bgprofile}
This section presents useful analytical expressions for the profiles we use in this work, including NFW \citep{Navarro97}, Burkert \citep{Burkert95}, and a profile that describes the stellar density of NGC5846-UDG1. 
The density profiles are already given in the main text, here we list the enclosed mass ($M$), gravitational potential ($\Phi$), gravitational acceleration in the cylindrical coordinate system ($f_R$, $f_\phi=0$, $f_z$), and the one-dimensional velocity dispersion ($\sigma$) for an isotropic velocity distribution are presented. 

\subsection{NFW}
\be
\label{eq:NFWmass}
    M(r) = \Mh \frac{f(x)}{f(c)},
\ee
where $x=r/\rs$ and $ f(x)=\ln(1+x) - x/(1+x)$.

\be
\label{eq:NFWpotential}
    \Phi(r) = \Phi_0 \frac{\ln(1+x)}{x},
\ee
where $\Phi_0 = -4 \pi G \rho_0 r_{\rm s}^2$

\be
\label{eq:NFWgrav}
    f_R = -\frac{\partial\Phi}{\partial R} = \Phi_0 \frac{f(x)}{x} \frac{R}{r^2}\,\,\,,
f_z = -\frac{\partial\Phi}{\partial z} = \Phi_0 \frac{f(x)}{x} \frac{z}{r^2},
\ee
where $r = \sqrt{R^2+z^2}$.

\bad
\label{eq:eq:NFWvdisp}
    \sigma^2(r)&= -\Phi_0x(1+x)^2\int_{x}^{\infty} \frac{f(x')}{x'^3(1+x')^2} \text{d}x'\\
    &\approx  V_{\rm max}^2 \left(\frac{1.439 x^{0.354}}{1+1.176x^{0.725}}\right)^2
\ead
where the second line is an approximation accurate to $1\%$ for $x=0.01$-$100$ \citep{Zentner03}.

\subsection{Burkert}\label{sec:Burkert}
\be
\label{eq:Burkertmass}
    M(r) = 2\pi \rho_0 r_{\rm s}^3 g(x),
\ee
where $\rho_0 = \Mh/\left[2\pi r_{\rm vir}^3 g(c)c^3\right]$, $ g(x)=0.5\ln(1+x^2)+\ln(1+x) - \arctan x$, and $x=r/r_{\rm s}$.

\bad
\label{eq:Burkertpotential}
    \Phi(r) &= \frac{\Phi_0}{4x}\left\{2\left(1+x\right)\left[\arctan\frac{1}{x}+\ln\left(1+x\right)\right]\right.\\
    &\left.+\left(1-x\right)\ln\left(1+x^2\right)-\pi\right\},
\ead
where $\Phi_0 = -4\pi G \rho_0 r_{\rm s}^2$.

\be
\label{eq:Burkertgrav}
    f_R = -\frac{\partial\Phi}{\partial R} = \Phi_0 \frac{f_0}{x^2} \frac{R}{r_{\rm s}}\,\,\,,
f_z = -\frac{\partial\Phi}{\partial z} = \Phi_0 \frac{f_0}{x^2} \frac{z}{r_{\rm s}},
\ee
where $r = \sqrt{R^2+z^2}$ and $f_0=2\arctan\frac{1}{x}+2f(x)+2\arctan x-\pi$.

\bad
\label{eq:eq:Burkertvdisp}
    \sigma^2(r)&=-\frac{\Phi_0}{2} \left(1+x\right)\left(1+x^2\right)\int_{x}^{\infty} \frac{f(x')}{x'^2(1+x')(1+x'^2)} \text{d}x'\\
    &\approx  V_{\rm max}^{0.299}  \frac{e^{x^{0.17}}}{1+0.286x^{0.797}}.
\ead

\subsection{UDG1 stellar profile}\label{app:UDG1}

\bad
\label{eq:UDG1-mass}
    M(r)&=\frac{\Ms}{2(9+2\sqrt{3}\pi)}\left[\sqrt{3}\pi+18f_3(x)\right.\\
    &\left.-6\sqrt{3}f_2(x)+9f_3(x)\right],
\ead
where $x=r/r_{\rm s}$ with $\rs$ a scale radius ($\rs=2\kpc$); and
\begin{align}
    f_1(x)&=\ln\frac{1-x+x^2}{(1+x)^2},\\
    f_2(x)&=\arctan\frac{1-2x}{\sqrt{3}},\\
    f_3(x)&=\frac{x}{1+x}.
\end{align}

\bad
\label{eq:UDG1-potential}
    \Phi(r)&=\frac{\sqrt{3}\Phi_0}{54x}\left[(1+6x)\pi \right.\\
    &\left.+6(2x-1)f_2(x)+3\sqrt{3}f_1(x)\right],
\ead
where $\Phi_0 = -4\pi G \rho_0 r_{\rm s}^2$.

\be
\label{eq:UDG1-grav}
    f_R= -\frac{\partial\Phi}{\partial R} = \frac{\Phi_0C}{54r^2}\frac{R}{x}\,\,\,,f_z = -\frac{\partial\Phi}{\partial z}= \frac{\Phi_0C}{54r^2} \frac{z}{x},
\ee
where $C=\sqrt{3}\pi+18f_3(x)-6\sqrt{3}f_2(x)+9f_1(x)$.

\bad
\label{eq:UDG1-vdisp}
    \sigma^2(r)&= -\frac{\Phi_0}{54}\left(1+x\right)\left(1+x^3\right)\\
    &\times\int_x^\infty\frac{\sqrt{3}\pi+18f_3(x')-6\sqrt{3}f_2(x')+9f_3(x')}{x'^2(1+x')^2(1-x'+x'^2)} \text{d}x'\\
    &\approx -\frac{\Phi_0}{54}\frac{1.845e^{x^{0.104}}}{1+0.563x^{1.158}}
\ead
where the second line is an approximation specifically for UDG1.

\section{GC profile: EFF profile}\label{sec: EFFprofile}
The EFF profile \citep{Elson87} is specified by the total mass, $m_{\rm tot}$, the scale length $a$, and the power-law slope $\eta$ -- the density profile is given by
\be\label{eq:EFFdensity}
\rho(l) = \frac{\rho_0}{(1+l^2/a^2)^\eta},
\ee
where 
\be
\rho_0 = \frac{\Gamma(\eta)}{\pi^{3/2}\Gamma(\eta-1)}\frac{m_{\rm tot}}{a^3}
\ee
is the central density, with $\Gamma(x)$ the Gamma function.

The enclosed mass of EFF profile is given by
\be
m(l) = \frac{4\pi}{3} l^3 \rho_0 \mathcal{F}_{21}\left(\frac{3}{2},\eta;\frac{5}{2};-\frac{l^2}{a^2}\right),
\ee
where $\mathcal{F}_{21}(a,b;c;z)$ is the hypergeometric function.
By solving $m(l) = 0.5 m_{\rm tot}$, one can show that the half-mass radius is given by
\be
\lhalf =(2^{\frac{2}{2\eta-3}}-1)^{1/2} a,
\ee
a quantity that is repeated used in our model.

As mentioned in \se{mass-size}, to estimate the tidal heating parameter $\ft$, we have we have used the tidal evolution track of \citet{Penarrubia10} expressed in terms of the maximum-circular velocity $\vmax$ and the radius $\lmax$ at which $\vmax$ is reached. 
To this end, we need a relation between $\lmax$ and the parameters defining the profile, which is obtained as follows.
The radius at which the circular velocity reaches maximum, $l_{\rm max}$, is given by the solution of $\dd v_{\rm circ}^2/\dd l =0$, i.e., 
\be
\mathcal{F}_{21}\left(\frac{3}{2},\eta;\frac{5}{2};-\frac{l^2}{a^2}\right)-\frac{3\eta}{5}\frac{l^2}{a^2}\mathcal{F}_{21}\left(\frac{5}{2},\eta+1;\frac{7}{2};-\frac{l^2}{a^2}\right) = 0,
\ee
and is well approximated by
\be\label{EFFrmax}
l_{\rm max} \approx 1.825 a
\ee
for $\eta=2$.

\section{GC mass-distance relations}\label{app:interpolation}
As mentioned in \se{MCMC}, to facilitate parameter inference, we pre-compute the model predictions of the observables used to construct the likelihood using one thousand GCs on a mesh grid spanned by the parameters of interest. 
We perform linear interpolation for model evaluations during the MCMC run. 

\begin{figure*}
\centering
\includegraphics[width=0.9\textwidth]{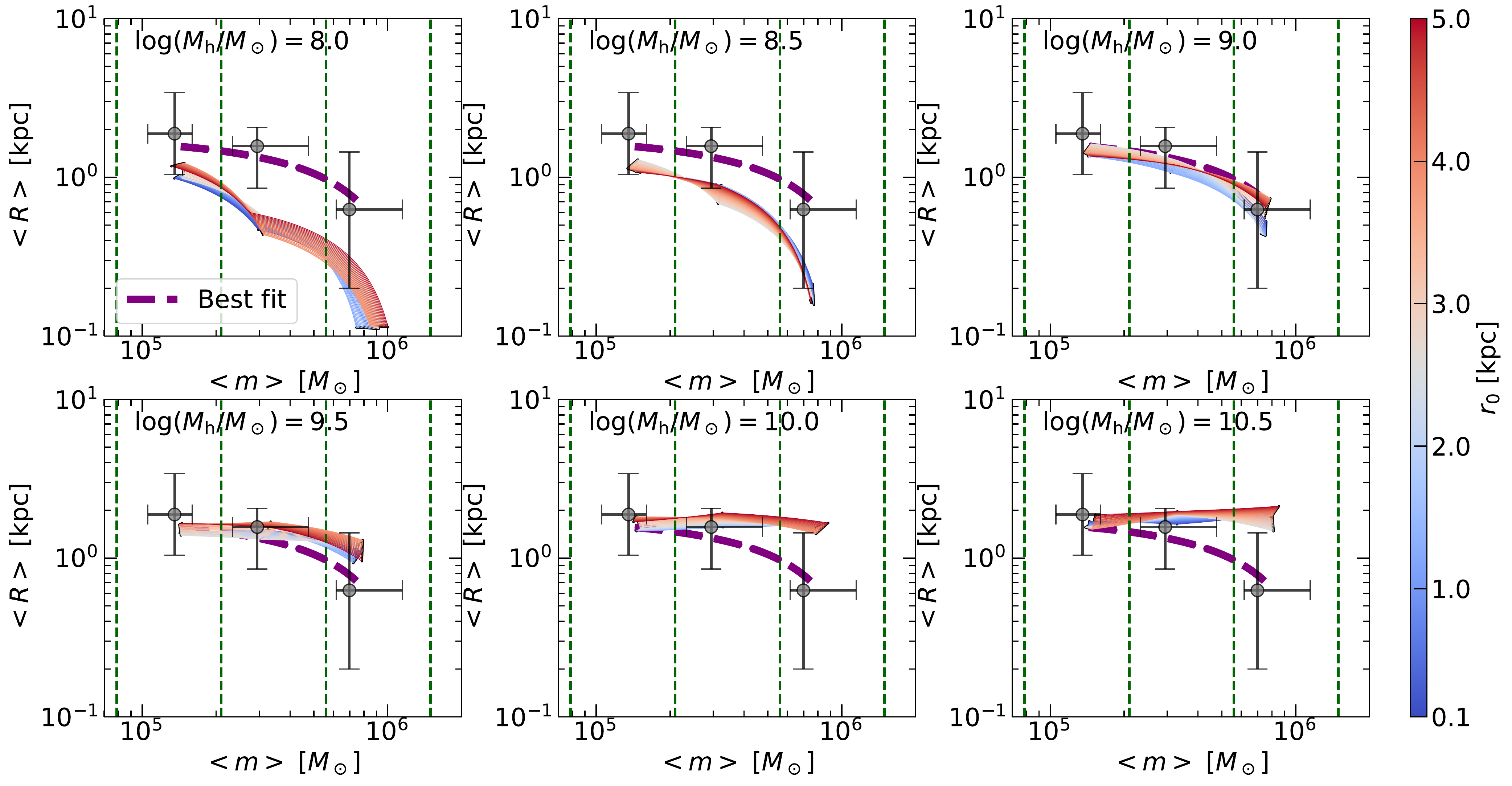}
\caption{The median galactocentric distance $\langle R\rangle$ versus the median GC mass $\langle m\rangle$, in three mass bins, for different host halo masses ($\Mh$) and different scale distance of the initial star-cluster distribution ($r_0$). The green dashed lines stand for the boundaries of the mass bins. The gray circles represent the data with the error bars indicating the 16th and 84th percentiles (the same across panels). The lines are model realizations for NFW halos with concentration $c=c_{-2}=4$ fixed. The purple dash lines stand for the interpolation with the best fit parameters}
    \label{fig:median_NFW}
\end{figure*}

\begin{figure*}
\centering
\includegraphics[width=0.9\textwidth]{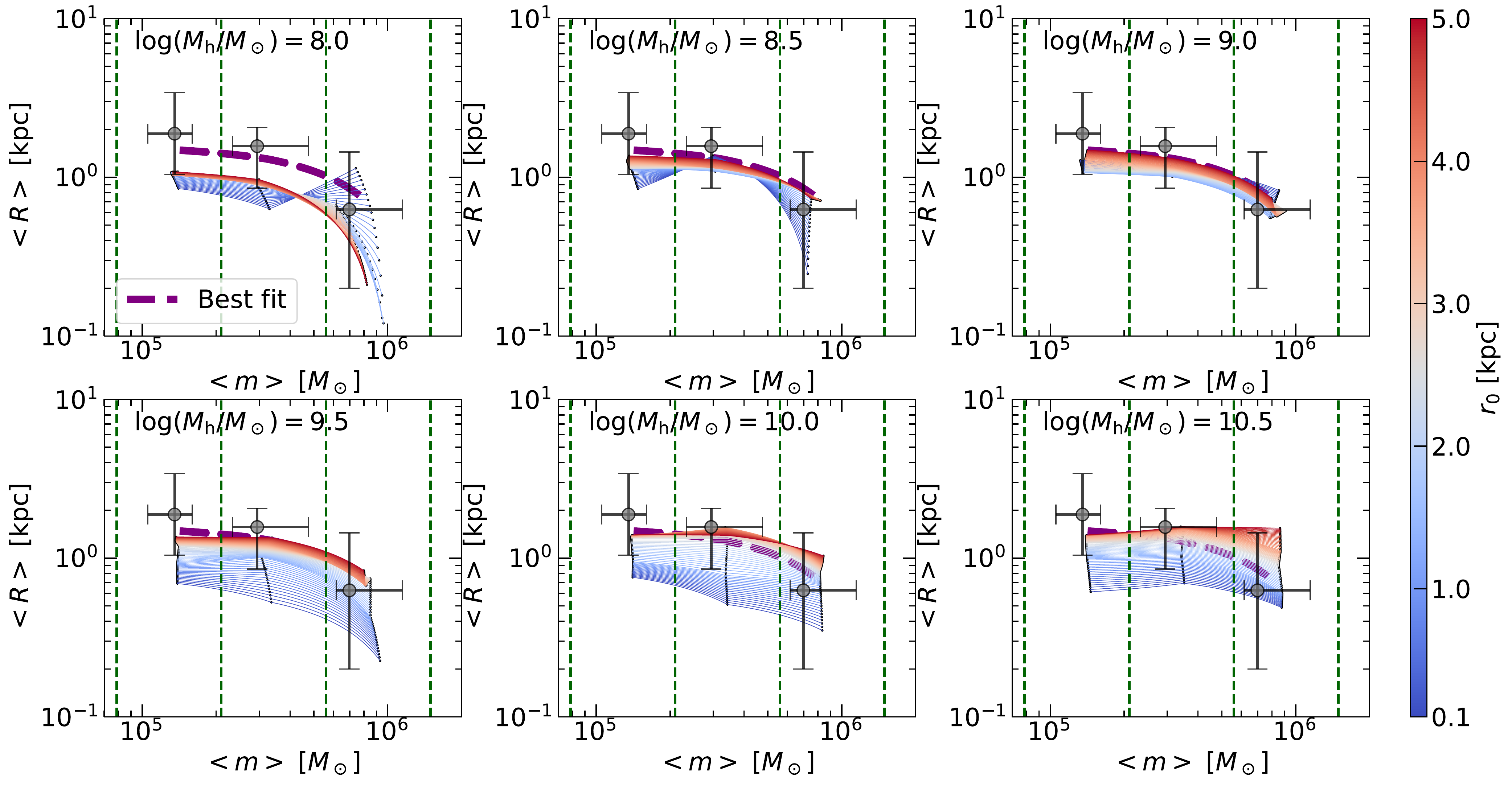}
\caption{The same as \Fig{median_NFW}, but for Burkert halos with $c=25$ ($c_{-2}=16$).}
\label{fig:median_Burkert}
\end{figure*}

Using the tabulated results, we plot the median relations between the evolved GC mass and distance for different halo mass $\Mh$ and different scale length $r_0$ of the initial star-cluster distribution, in \Figs{median_NFW}-\ref{fig:median_Burkert}.
The concentrations are kept fixed at the posterior mode values, i.e., \Fig{median_NFW} is for NFW halos of $c=c_{-2}=4$ and \Fig{median_Burkert} is for Burkert halos of $c=25$ ($c_{-2}=16$). 
For the NFW case, we can see clear mass segregation at $\Mh \la 10^{9.5}\Msun$, not very sensitive to $r_0$. 
For the Burkert case, mass segregation is achieved at $\Mh \la 10^{10}\Msun$ and but requires a relatively large $r_0$. 
In both cases, the halo mass cannot be too small ($\la 10^{8.5}\Msun$), otherwise the median distances would be too small. 

\section{Effect from initial mass segregation}\label{app:init_segre}

Throughout this study, we have assumed that there is no mass segregation in the initialization. However, there might be mass segregation $\sim$10 Gyr ago if the molecular clouds which form GCs later had experienced dynamical friction already; or it might be that the gas-disk instability imprinted some primordial mass segregation. 
Here we explore the effect of such an initial mass segregation. 
In particular, since our best-fit halo mass is low, we aim to qualitatively answer the question -- what level of primordial mass segregation can give rise to the final observed segregation strength, in a halo that is significantly more massive that our posterior mode masses and thus yields a much weaker DF effect. 

\begin{figure*}
\centering
\includegraphics[width=0.5\textwidth]{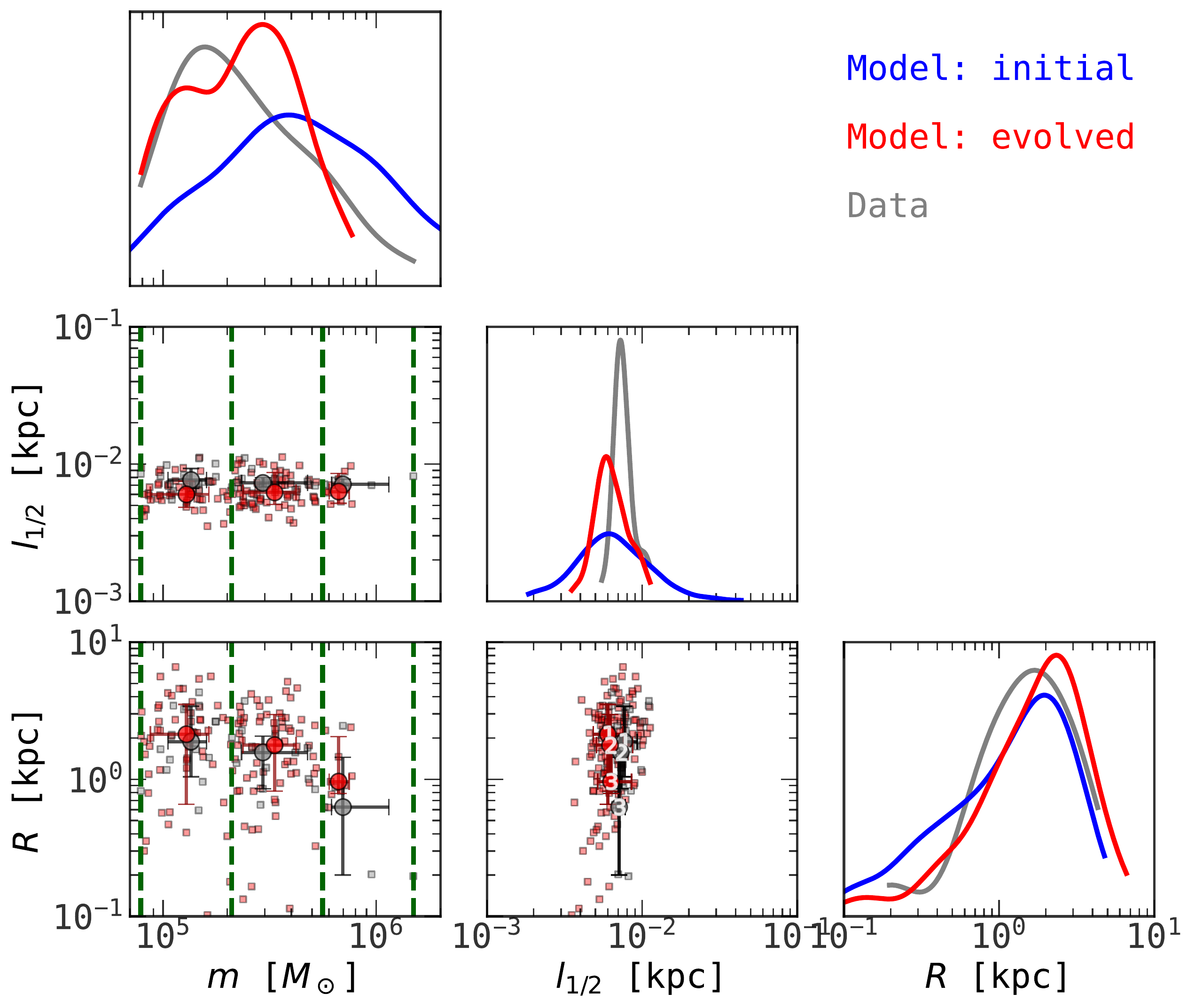}
\caption{The same as \Fig{NFWrealization} and \Fig{Burkert-realization} but with an high-mass NFW host halo ($\Mh=10^{9.8}\Msun$, $c=15$). 300 GCs are initialized in three mass groups with equal number, according to different Hernquist profiles such that the initial slope is -0.52.}
\label{fig:goodinit}
\end{figure*}

\begin{figure*}
\centering
\includegraphics[width=0.5\textwidth]{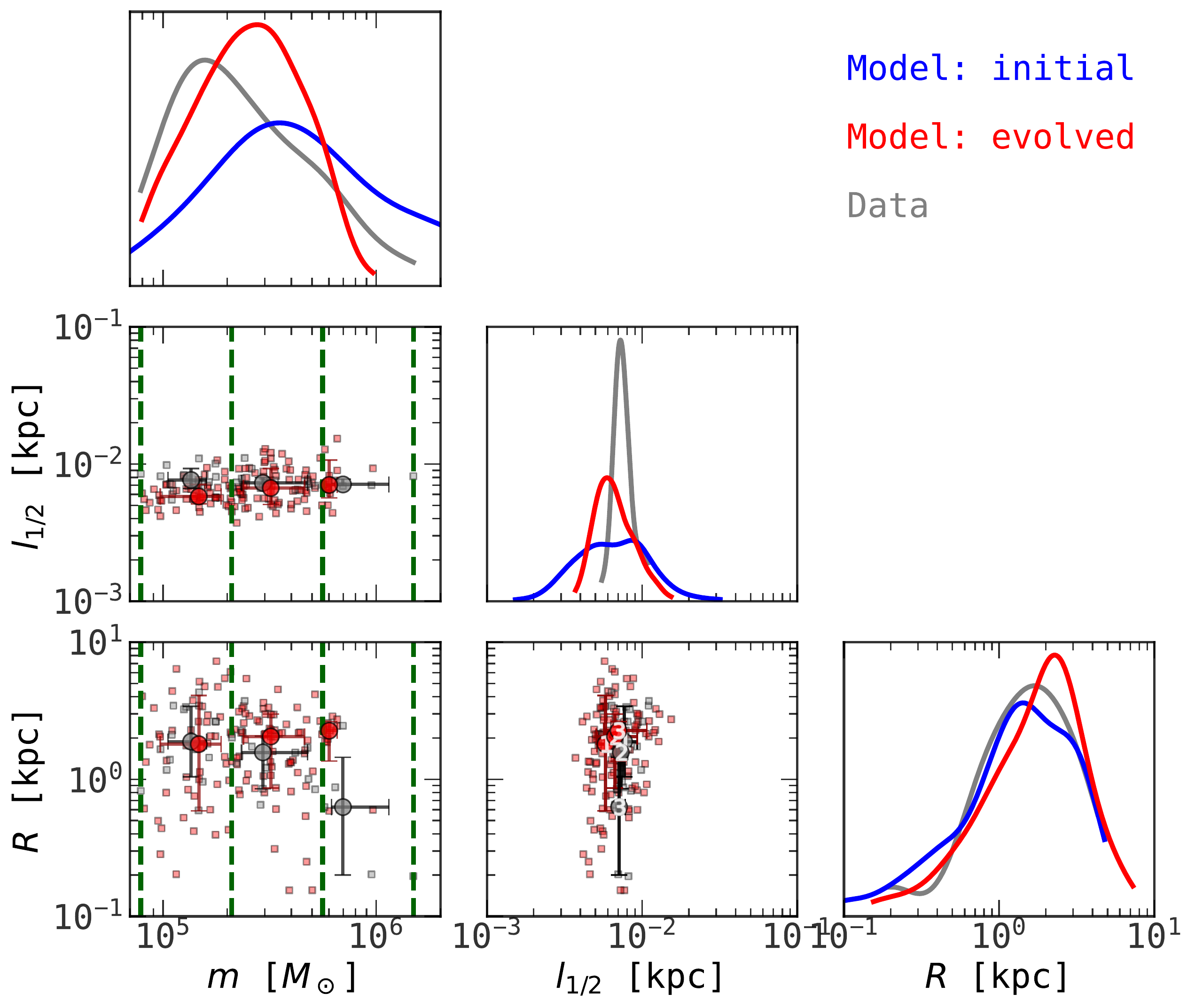}
\caption{The same as \Fig{goodinit}, but with a shallower initial mass-segregation slope of -0.35.}
\label{fig:badinit}
\end{figure*}

In \Figs{goodinit} and \ref{fig:badinit}, we show the model results assuming initial mass segregation of different slopes. 
We keep all the other parameters the same as before, i.e., $\Mmin=10^{5.5}\Msun$ and $\Mmax=10^8\Msun$ for the ICMF and draw 300 GCs. 
To obtain an initial segregation, instead of drawing all the GCs from a single Hernquist profile, we split them into three mass bins of 100 GCs each. We draw the distances of the GCs in each group with a different Hernquist profile. 
For the low-mass GCs, we adopt a scale radius of $r_{0,\rm low}=4.8$ kpc. 
To get different initial mass-segregation levels, we adopt $r_{0,\rm mid}=2.8$ kpc  and 3.5 kpc for the intermediate-mass GCs, for the experiments shown in \Fig{goodinit} and \Fig{badinit}, respectively; and, similarly, $r_{0,\rm high}=0.4$ kpc and 1.5 kpc, respectively. 
We also limit the initial radius range of the three groups by choosing different minimal radius $r_{\rm min}$ and maximal radius $r_{\rm max}$ -- for low-mass GCs, we set $r_{\rm min,low}=1$ kpc and $r_{\rm max,low}=5$ kpc; for intermediate-mass GCs, we set $r_{\rm min,low}=0.5$ kpc and $r_{\rm max,low}=4$ kpc; and for high-mass GCs, we use $r_{\rm min,low}=0.1$ kpc and $r_{\rm max,low}=2$ kpc. 
Such initialized, the slope of the initial mass segregation is -0.52 in \Fig{goodinit}, and is -0.35 in \Fig{badinit}, comparable to and much smaller than the observed slope of -0.57, respectively. 
Then we evolve them in a NFW halo of $10^{9.8} M_\odot$ and $c=15$, which will not produce significant mass segregation with our fiducial setup.
Not surprisingly, in such a relatively massive, normal-concentration halo, an initial mass segregation that is marginally weaker than the final one can basically reproduce the final mass segregation; whereas a significantly weaker initial mass segregation cannot.

\end{document}